\documentclass[useAMS,usenatbib]{mn2e}
\usepackage{graphicx}
\usepackage{hyperref}

\newcommand{\Os}{\Omega_s}
\newcommand{\Oc}{\Omega_c}
\newcommand{\DO}{\Delta\Omega}
\newcommand{\DOC}{{\Delta\Omega_{\rm cr}}}
\newcommand{\DOCM}{{{\Delta\Omega_{\rm cr}}_{\rm max}}}

\newcommand{\Od}{\dot{\Omega}}
\newcommand{\DTG}{\Delta t_{\rm gl}}
\newcommand{\xm}{x_{\rm max}}
\newcommand{\Nv}{N_{\rm  v}}
\newcommand{\DLg}{\Delta L_{\rm gl}}
\newcommand{\Ygl}{Y_{\rm gl}}
\newcommand{\DOgl}{\Delta\Omega_{\rm gl}}
\newcommand{\fpm}{{f_{\rm PM}}}
\newcommand{\Acc}{{\Delta\dot{\Omega}_{gl}/\dot{\Omega}_\infty}}
\newcommand{\ACC}{\frac{{\Delta\dot{\Omega}_{gl}}}{\dot{\Omega}_\infty}}
\newcommand{\xRat}{\xm/R_{\rm ic}}

\title[Realistic EOSs and the ``snowplow'' glitch model.]{The effect of realistic equations of state and general relativity on the ``snowplow'' model for pulsar glitches.}
\author[Seveso et al.]{S.~Seveso$^{1,2}$, P.~M.~Pizzochero,$^{1,2}$\thanks{E-mail:pierre.pizzochero@mi.infn.it}, B.~Haskell$^{3}$\\
$^{1}$Dipartimento di Fisica, Universit\`a degli Studi di Milano, Via Celoria 16, 20133 Milano, Italy\\
$^{2}$Istituto Nazionale di Fisica Nucleare, sezione di Milano, Via Celoria 16, 20133 Milano, Italy\\
$^{3}$Astronomical Institute ``Anton Pannekoek'', University of Amsterdam, Postbus 94249, 1090 GE Amsterdam, the Netherlands}

\begin{document}


\pagerange{\pageref{firstpage}--\pageref{lastpage}} \pubyear{2012}

\maketitle

\label{firstpage}

\begin{abstract}

Many pulsars are observed to ``glitch'', i.e. show sudden jumps in their rotational frequency $\nu$, some of which can be as large as $\Delta \nu/\nu\approx 10^{-6}-10^{-5}$ in a subset of pulsars known as giant glitchers. Recently Pizzochero (2011) has shown that an analytic model based on realistic values for the pinning forces in the crust and for the angular momentum transfer in the star can describe the average properties of giant glitches, such as the inter-glitch waiting time, the step in frequency and that in frequency derivative. In this paper we extend the model (originally developed in Newtonian gravity and for a polytropic equation of state) to realistic backgrounds obtained by integrating the relativistic equations of stellar structure and using physically motivated equations of state to describe matter in the neutron star. We find that this more detailed treatment still reproduces the main features of giant glitches in the Vela pulsar and allows us to set constraints on the equation of state. In particular we find that stiffer equations of state are favoured and that it is unlikely that the Vela pulsar has a high mass (larger than $M\approx 1.5 M_\odot$).   

\end{abstract}

\begin{keywords}
stars:neutron - pulsars:general - pulsars:individual:PSR B0833-45
\end{keywords}

\section{Introduction}
Many pulsars are observed to ``glitch'', i.e. they show sudden increases in their spin frequency that are instantaneous to the accuracy of the data. To date several hundreds of glitches have been detected \citep{Espinoza}, with relative increases in the spin frequency $\nu$ that range from as low as $\Delta\nu/\nu\approx 10^{-11}$ to $\Delta\nu/\nu\approx 10^{-5}$. In particular a class of pulsars, of which the Vela pulsar is the prototype, exhibit what are known as ``giant'' glitches \citep{Espinoza}, large steps in the spin frequency ($\Delta\nu/\nu\approx 10^{-6}$) which are accompanied by an increase in the spindown rate $\dot{\nu}$ and exhibit a rough periodicity in their recurrence rate (for example giant glitches in the Vela occur roughly every three years). 

Shortly after the first glitches were observed it was suggested that they could be due to a superfluid component in the stellar interior, weakly coupled to the normal component and to the electromagnetic emission, that could store angular momentum and then release it catastrophically, giving rise to a glitch \citep{Baym,AndItoh75,Alpar77,Alpar84}. Large scale superfluid components are, in fact,  expected in Neutron Star (NS) interiors on theoretical grounds given that the temperature of the star will drop below the superfluid critical temperature (typically $\approx 10^9$ K) soon after birth. Furthermore recent observations of the cooling of the young NS in the supernova remnant Cassiopeia A have provided the first direct indication of superfluidity in NS interiors \citep{CASA1,CASA2}. 

A superfluid rotates by forming an array of quantised vortices which carry the circulation of the fluid. In the NS crust the vortices can be strongly attracted, ``pinned'', to the nuclear lattice \citep{AndItoh75,Alpar77,Pines80,Anderson82} and cannot move outward. If the superfluid cannot remove vortices it cannot spin down and it therefore acts as an angular momentum reservoir. As the crust spins down due to electromagnetic emission a lag will develop between the superfluid and the normal component, leading to a hydrodynamical lift force (Magnus force) acting on the vortices. Eventually when the lag reaches a critical value the pinning force will no longer be able to contrast the hydrodynamical lift and the vortices will unpin, transferring their angular momentum to the crust and giving rise to a glitch. 

Although there is some evidence that smaller glitches in young active pulsars such as the Crab may be related to crust quakes \citep{Crow,Mid06}, there is a growing consensus that the basic picture outlined above can be used to describe the main features of pulsar glitches. There is still considerable debate on the ``trigger'' for vortex unpinning and several mechanisms have been proposed (including crust quakes \citep{Rud76,Rud}, heat release \citep{LL} and hydrodynamical instabilities \citep{kglitch}), but recent work by \citet{Melatos1} (see also \citet{Melatos2,Melatos3}) has been very successful in using cellular automaton simulations that can track the movement of a large number of vortices, to reproduce the distribution of glitch sizes and waiting times, and \citet{Hask12} have produced the first hydrodynamical simulation that can follow all stages of a giant glitch, from the rise to the relaxation.

One of the main difficulties in performing such calculations, up to now, has been the relative scarcity of realistic estimates of the pinning force between vortices and nuclei in the crust. However the recent calculation of \citet{Gpaper1} (see also \citet{Gtesi,Gpaper2}) has filled this gap and produced physically consistent pinning profiles that can be used to study pulsar glitches. In fact \citet{Pizzochero} has shown that these forces can be incorporated into a simple analytical model, the so-called ``snowplow'' model, that predicts the size, step in frequency derivative and waiting time of Vela glitches. The same forces were used by \citet{Hask12} in a hydrodynamical model to reproduce also the post-glitch relaxation of Vela glitches and show that the model is consistent with the size and waiting times of other pulsars that exhibit giant glitches.

We shall discuss the details of the snowplow model in the following sections, the main assumption, however, is that vortices close to the rotational axis of the star will only be weakly pinned at their extremities. The Magnus force can thus easily unpin them and move them towards the equator, where they will repin as they are now immersed in the strong pinning region of the inner crust. This creates a vortex sheet that moves close to the maximum of the pinning potential as the star spins down. Once the maximum critical lag is reached the pinning force can no longer balance the Magnus force, and all the vorticity that has been accumulated is free to go, giving rise to a giant glitch. This also gives rise to a natural periodicity for giant glitches, although crust quakes or vortex avalanches can unpin part of the vorticity and give rise to smaller glitches before the maximum.

In this picture we implicitly assume that vortices that cross the core of the star are immersed in a low drag environment, i.e. that there is no pinning in the core and vortices are free to move out. This would not be the case if the protons in the interior are in a type II superconducting state and there is a strong interaction between magnetic flux tubes and rotational vortices \citep{Rud,Link03}. Note, however, that a large portion of the star may be in a type I superconducting state \citep{Jones06} in which the magnetic field is not organised in flux tubes, but rather in macroscopic regions of normal matter, and the interactions may be much weaker \citep{Sedrakian05} (although see \citet{Jones06} for a discussion of strong interactions in type I superconductors). Furthermore recent calculations suggest that even if the superconductivity is of type II, the interaction between vortices and flux tubes will be weak in the presence of strong entrainment or superfluid $\Sigma^{-}$ hyperons \citep{Babaev}. In this paper we thus take the view that pinning in the core will be weak, although strong pinning of vortices to flux tubes is an intriguing possibility and will be the focus of a future publication (Haskell, Pizzochero \& Seveso, in preparation).

\citet{Pizzochero} and \citet{Hask12} developed the model for an $n=1$ polytrope in Newtonian gravity, in order to obtain simple analytical estimates for the glitch size. In this paper we investigate the effect on the snowplow model of \citet{Pizzochero} of using realistic Equations of State (EoSs) and relativistic equations for the stellar structure. We will see that in general the model is consistent with observations and that less massive NSs (in the region of $1.2 M_\odot-1.5 M_\odot$) are favoured to describe the Vela pulsar.
The inclusion of realistic backgrounds and equations of state in the hydrodynamical models will be the focus of future work.

\section[]{Stellar structure}

\begin{figure}
	\centering
	\includegraphics{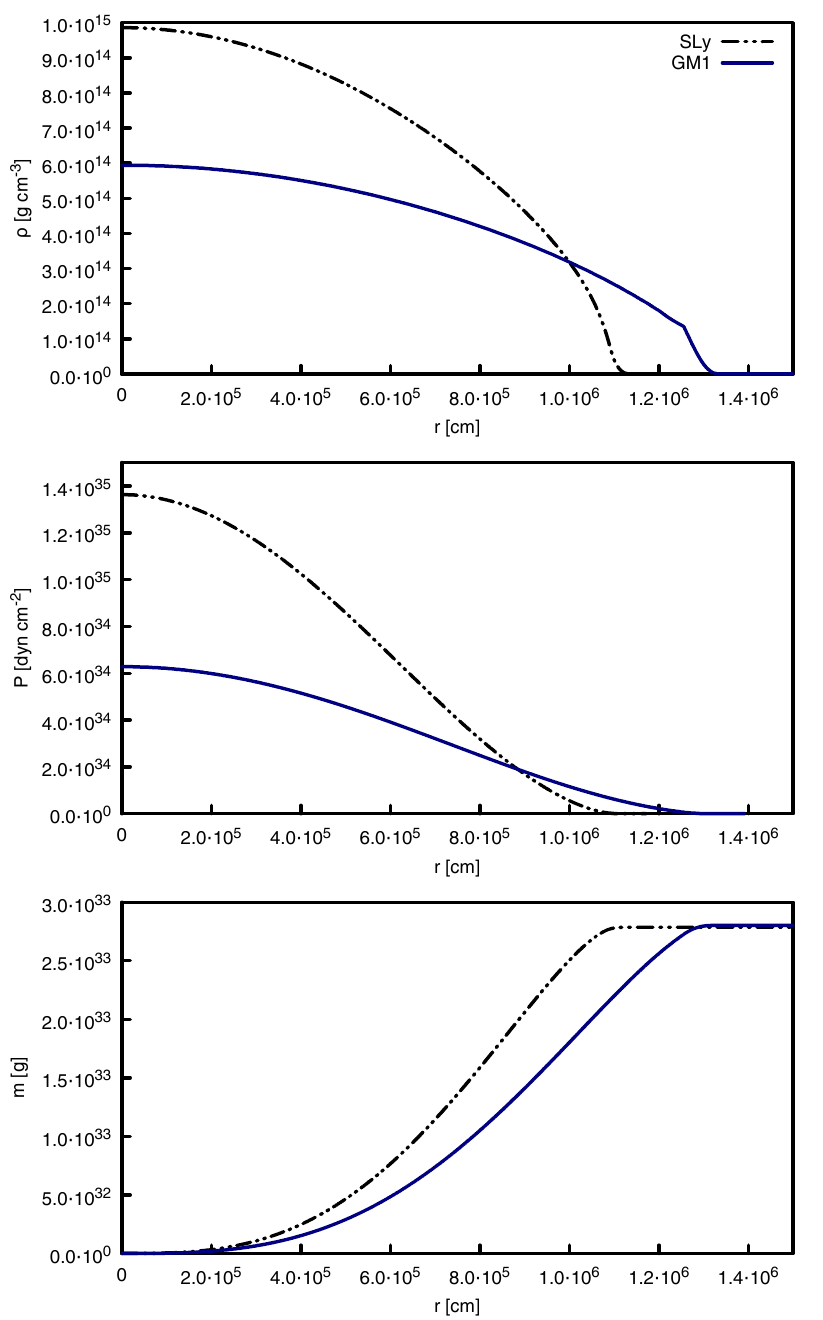}
	\caption{Mass, pressure and density profiles of 1.4 $M_\odot$ neutron star for the EoSs considered: with fixed mass, a stiffer EoS produce a star with larger radius and lower central density.}
	\label{fig:tov}
\end{figure}

As outlined above the model proposed by \citet{Pizzochero} relies on calculating the amount of vorticity that can be stored in the inner crust before the Magnus force overcomes the maximum pinning force and then calculating the amount of angular momentum exchanged between the superfluid component and the crust.
In order to obtain realistic estimates of the moment of inertia of the different components and of the strength of the density-dependent pinning force per unit length, it is thus important to use a realistic equation of state (EoS) for dense matter that describes the relation $P(\rho)$ between density and pressure. The stellar structure and density profile can then be obtained by integration of the Tolman-Oppenheimer-Volkov (TOV) equations: 
\begin{eqnarray}
	\frac{dm(r)}{dr} & = & 4\pi r^2 \rho(r) \\
	\frac{d\phi(r)}{dr} & = & \left( \frac{Gm(r)}{r^2} +4\pi G r\,\frac{P(r)}{c^2} \right) \left( 1 - \frac{2Gm(r)}{c^2r} \right)^ {-1}\\
	\frac{dP(r)}{dr} & = & -\left(\rho(r) + \frac{P(r)}{c^2} \right) \, \frac{d\phi}{dr},
\label{eqn:TOV}
\end{eqnarray} 
where $m(r)$ is the mass contained in a sphere of radius $r$, $\rho(r)$ is the density profile and $P(r)$ is the pressure. These differential equations model the hydrostatic equilibrium inside the star with relativistic approach and, of course, require the $P(\rho(r))$ function. The last two equations can be combined in one that gives an expression for the mass and pressure derivatives and the system can be solved with valid initial conditions. We obtain the functions that describe the star with the fourth--order Runge--Kutta method, starting at $r=0$ with $m(0) = 0$ and $\rho(0) = \rho_c$, for a valid choice of the central density $\rho_c$. The integration stops when we reach the condition $\rho(R) = 10^{-8} \rho_c$ and we take $R$ as the radius of the star. Of course the mass of the star is $M = m(R)$. As a result of this integration we have $m(r$), $\rho(r)$ and $P(r)$ of the star as showed in Fig. \ref{fig:tov}.

\begin{figure}
	\centering
	\includegraphics{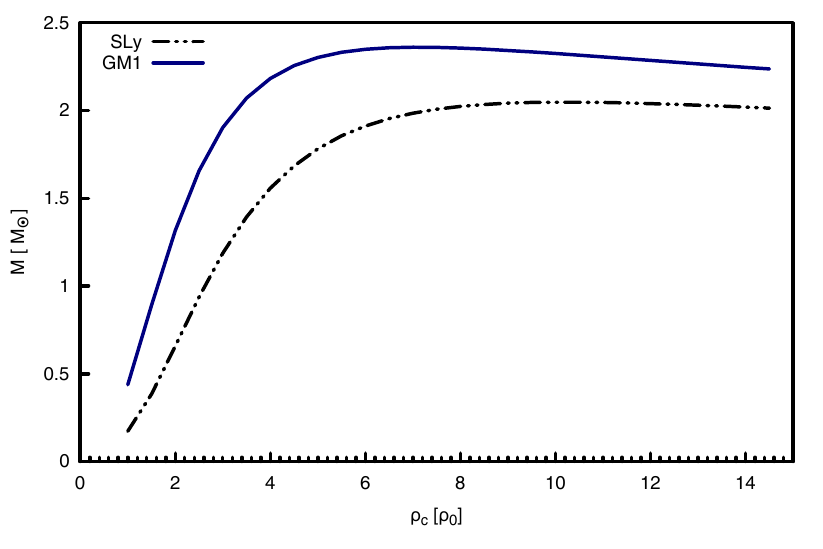}
	\caption{This plot shows the mass--central density relation for the EoSs considered. As expected, we find  a maximum mass value above $2 M_\odot$ for each equation of state (see table \ref{tab:maxmass}).}
	\label{fig:maxmass}
\end{figure}

\noindent We use two different EoSs: 
\begin{enumerate}
	\item SLy \citep{DH01} is a moderate EoS, based on a non--relativistic parametrisation; this equation describes the whole star with a single analytical expression and so it is more convenient to integrate;
	\item GM1 by \citet{GM1} is a stiff $P(\rho)$ relation that is very similar to SLy in the crust of star, but not in the core due to different microscopic approach used to describe hadrons at densities higher than $\rho_0$.
\end{enumerate}
Fig. \ref{fig:maxmass} shows the different mass--central density relations for the two equations of state.

This figure clearly identifies also the presence of a limit for the mass for a neutron star. The existence of a maximum mass $M_{\rm max}$ is an effect of the relativistic nature of the TOV equations, where pressure contributes to the gravitational field. Of course $M_{\rm max}$ depends on the EoS: a stiffer equation of state gives a higher $M_{\rm max}$. The evidence of the existence of a $2M_\odot$ neutron star \citep{Dem} allows, in fact, to reject soft equations of state that predict a maximum mass below this value. Table \ref{tab:maxmass} shows the different values of $M_{\rm max}$ for the equations of state considered.

\begin{table}
\centering
\caption{This table shows, for each EoS, the maximum allowed mass with the corresponding central density $\rho_c$ (in units of nuclear saturation density $\rho_0$), radius of the star $R$, radius of the core $R_c$ and radius of the inner crust $R_{ic}$.}
\label{tab:maxmass}
\setlength{\tabcolsep}{12pt}
\begin{tabular}{@{}lrrrrr@{}}
\hline
{\bf EoS} & $\rho_c$ & $M$ & $R$ & $R_c$ & $R_{\rm ic}$ \\
 & ($\rho_0$) & ($M_\odot$) & (km) & (km) & (km) \\
\hline
SLy &   10.2 &   2.05 &   9.98 &   9.68 &   9.86 \\
GM1 &    7.1 &   2.36 &  11.98 &  11.57 &  11.82 \\
\hline
\end{tabular}
\end{table}

In this work we consider stars with masses from $1M_\odot$ to $M_{\rm max}$. With the density profile $\rho(r)$ it is possible to identify, for each star, the structural regions that are relevant for the model and important for the pinning. In particular we calculate the radius of the core $R_c$ as the distance from the center of the star where $\rho_{\rm core} = \rho(R_c) = 0.5\rho_0$ ($\rho_0 = 2.8 \times 10^{14} \,\mbox{g}\,\mbox{cm}^{-3}$ is the nuclear saturation density); the inner crust--outer crust interface $R_{\rm ic}$ corresponds, on the other hand, to the density value $\rho_d = 0.0015\rho_0$ that is the neutron drip point: this means than in the outer crust there are no free neutrons. It is easy to calculate the moment of inertia of a shell delimited by radii $r_1$ and $r_2$:
\begin{equation}
I(r_1, r_2) = \frac{8\pi}{3} \int_{r_1}^{r_2} r^4\rho(r) \,dr.
\label{eq:inertia}
\end{equation} 
We can then calculate also the moment of inertia of every region, considering that $I_{\rm core} = I(0, R_c)$, $I_{\rm ic} = I(R_c, R_{\rm ic})$ and $I_{\rm oc} = I(R_{\rm ic}, R)$.

Table \ref{tab:stars} shows all the relevant parameters for the considered stars, obtained from the integration of the TOV equations with SLy and GM1.

\begin{table*}
 \centering
 \begin{minipage}{170mm}
 \caption{We give all the structural parameters (as defined in section 2) of the stars used to test the snowplow model, for both EoSs tested. See also figure \ref{fig:thickness} for a graphical representation of these quantities.}
 \label{tab:stars}
 \setlength{\tabcolsep}{12pt}
\begin{tabular}{@{}lrrrrrrrr@{}}
\hline
{\bf EoS} & $M$ & $R$ & $R_c$ & $R_{\rm ic}$ & $I_{\rm tot}$ & $I_{\rm core}$ & $I_{\rm ic}$ & $I_{\rm oc}$ \\
 & ($M_\odot$) & (km) & (km) & (km) & ($10^{45}$ g$\,$cm$^2$) & ($10^{45}$ g$\,$cm$^2$) & ($10^{43}$ g$\,$cm$^2$) & ($10^{40}$ g$\,$cm$^2$)\\
\hline
{\bf SLy} &    1.0 &  11.86 &  10.35 &  11.23 &  0.739 &  0.697 &  4.181 &  6.638 \\
{\bf    } &    1.1 &  11.83 &  10.49 &  11.28 &  0.827 &  0.788 &  3.923 &  5.945 \\
{\bf    } &    1.2 &  11.80 &  10.60 &  11.31 &  0.914 &  0.878 &  3.652 &  5.317 \\
{\bf    } &    1.3 &  11.76 &  10.69 &  11.32 &  0.999 &  0.965 &  3.370 &  4.738 \\
{\bf    } &    1.4 &  11.71 &  10.75 &  11.32 &  1.079 &  1.048 &  3.078 &  4.198 \\
{\bf    } &    1.5 &  11.64 &  10.79 &  11.29 &  1.154 &  1.126 &  2.777 &  3.685 \\
{\bf    } &    1.6 &  11.55 &  10.79 &  11.24 &  1.222 &  1.197 &  2.469 &  3.194 \\
{\bf    } &    1.7 &  11.42 &  10.76 &  11.16 &  1.279 &  1.258 &  2.150 &  2.718 \\
{\bf    } &    1.8 &  11.26 &  10.68 &  11.03 &  1.322 &  1.303 &  1.818 &  2.248 \\
{\bf    } &    1.9 &  11.03 &  10.54 &  10.83 &  1.339 &  1.324 &  1.463 &  1.769 \\
{\bf    } &    2.0 &  10.62 &  10.23 &  10.47 &  1.299 &  1.289 &  1.042 &  1.233 \\
\hline
{\bf GM1} &    1.0 &  13.94 &  11.79 &  13.02 &  1.021 &  0.896 & 12.505 & 19.061 \\
{\bf    } &    1.1 &  13.94 &  12.01 &  13.12 &  1.146 &  1.025 & 12.068 & 17.532 \\
{\bf    } &    1.2 &  13.94 &  12.19 &  13.20 &  1.271 &  1.156 & 11.555 & 16.108 \\
{\bf    } &    1.3 &  13.93 &  12.35 &  13.27 &  1.395 &  1.285 & 10.991 & 14.780 \\
{\bf    } &    1.4 &  13.91 &  12.47 &  13.32 &  1.516 &  1.412 & 10.382 & 13.530 \\
{\bf    } &    1.5 &  13.89 &  12.58 &  13.34 &  1.634 &  1.536 &  9.738 & 12.340 \\
{\bf    } &    1.6 &  13.85 &  12.66 &  13.35 &  1.747 &  1.657 &  9.062 & 11.198 \\
{\bf    } &    1.7 &  13.79 &  12.71 &  13.35 &  1.854 &  1.771 &  8.362 & 10.099 \\
{\bf    } &    1.8 &  13.72 &  12.74 &  13.32 &  1.954 &  1.878 &  7.635 &  9.031 \\
{\bf    } &    1.9 &  13.62 &  12.74 &  13.26 &  2.043 &  1.974 &  6.885 &  7.987 \\
{\bf    } &    2.0 &  13.49 &  12.70 &  13.17 &  2.118 &  2.057 &  6.107 &  6.956 \\
{\bf    } &    2.1 &  13.33 &  12.63 &  13.05 &  2.173 &  2.120 &  5.292 &  5.922 \\
{\bf    } &    2.2 &  13.10 &  12.48 &  12.85 &  2.194 &  2.150 &  4.411 &  4.851 \\
{\bf    } &    2.3 &  12.71 &  12.20 &  12.51 &  2.146 &  2.113 &  3.371 &  3.631 \\
\hline
\end{tabular}
 \end{minipage}
\end{table*}

\begin{figure*}
	\centering
	\includegraphics{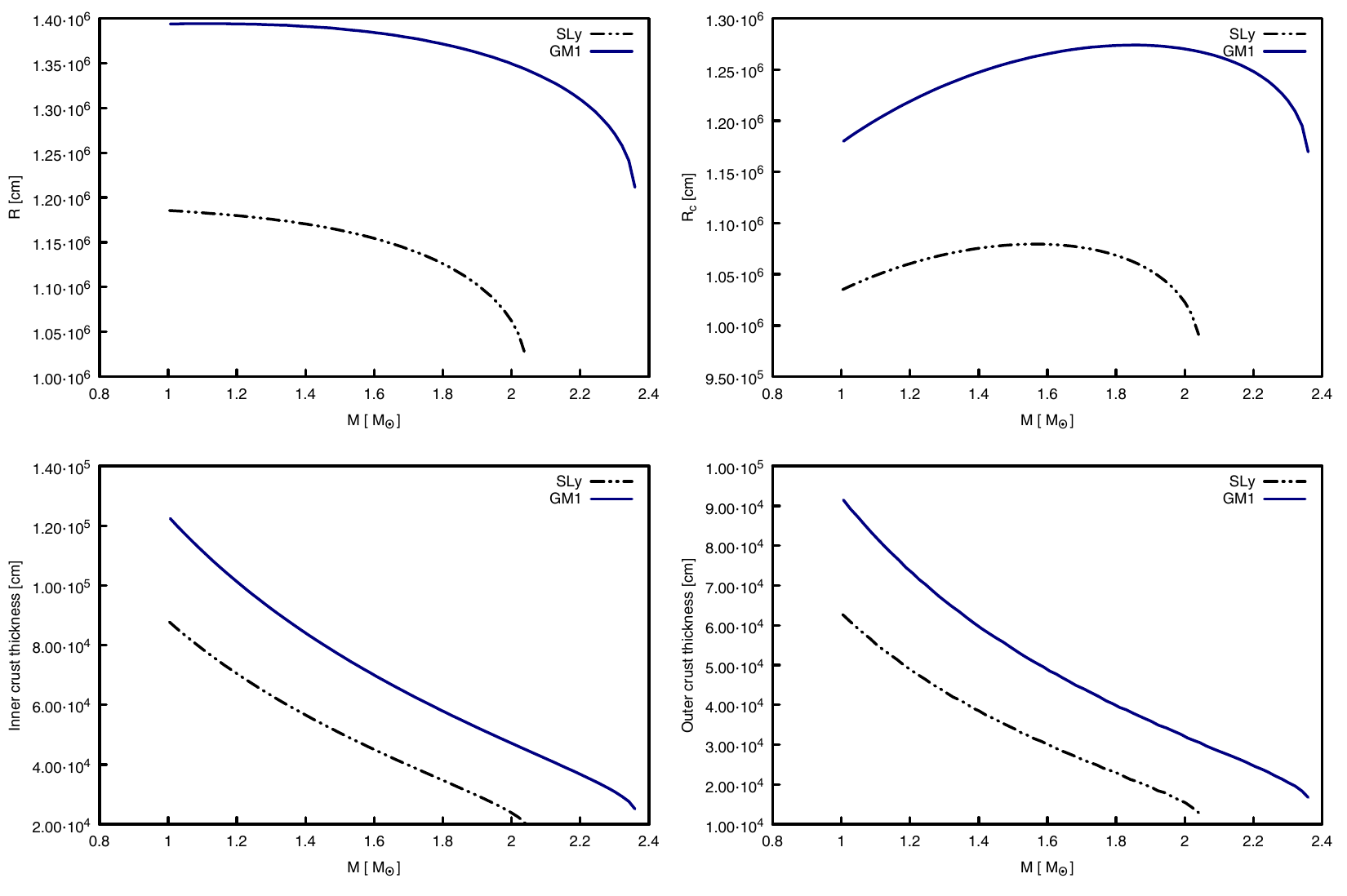}
	\caption{The first figure shows the dependence of the radius of the neutron star on the total mass, for the SLy and GM1 EoSs. The other plots represent the thicknesses of the stellar regions (core, inner crust and outer crust) as function of mass. As one can see a more massive star has thinner crusts, while a stiffer equation of state produces a larger star.}
	\label{fig:thickness}
\end{figure*}

\section{Pinning and vorticity}

One of the most important ingredients of the model is clearly  $f_{\rm pin}(\rho)$, the pinning force per unit length that acts on the vortex line as a result of its interaction with the lattice (in the inner crust). 
Although the pinning force per pinning site can readily be evaluated from the knowledge of the pinning energy \citep{Alpar77,EB,DP03,DP04,DP06}, the force per unit length of a vortex, which is the quantity that must be equated to the Magnus force in order to understand whether a vortex is pinned or free, is much more complex to evaluate, as it depends on the rigidity of a vortex and on its orientation with respect to the crustal lattice.

\begin{figure}
	\centering
	\includegraphics{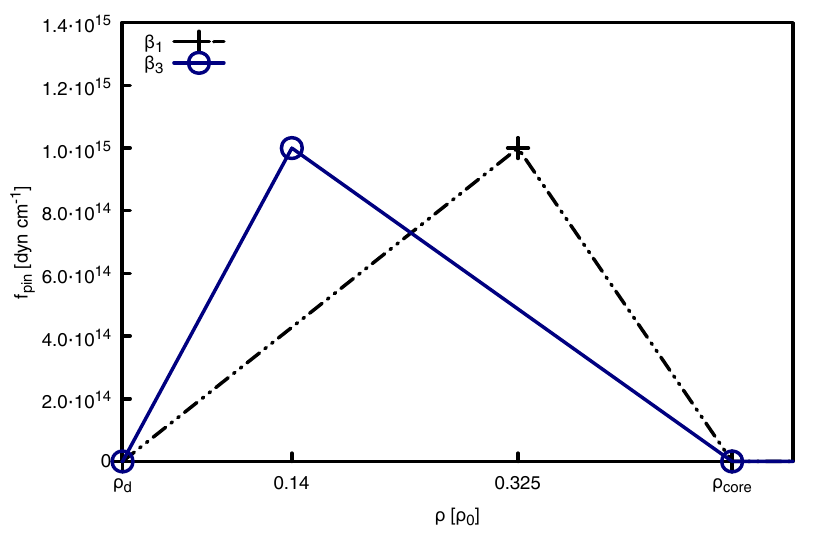}
	\caption{The profile of the pinning force $f_{\rm pin}(\rho)$ for the two cases $\beta = 1$ and $\beta = 3$, with a choice for the maximum of $\fpm = 10^{15}\,\mbox{dyn}\,\mbox{cm}^{-1}$.}
	\label{fig:pinning}
\end{figure}

\citet{Gpaper1} (see also \citet{Gtesi,Gpaper2}) have performed numerical simulations to evaluate this quantity, taking into account different orientation of the bcc lattice. They found that the order of magnitude of the maximum pinning force $\fpm$ is approximately $10^{15}\,\mbox{dyn}\,\mbox{cm}^{-1}$ and that there is no significant difference for the pinning force per unit length in considering vortex--nucleus interaction attractive or repulsive in different regions.
Another interesting result found by \citet{Gpaper1} regards the position of the maximum $\fpm$. In this paper the authors use a density--dependent pairing gap $\Delta(\rho)$ obtained with a realistic microscopic nucleon--nucleon interaction. It is known that the polarization effects of the neutron medium reduce the paring gap, but there is yet no agreement on how strong this suppression will be, although it seems reasonable to divide the $\Delta(\rho)$ by a factor $\beta$ between 2 and 3. \citet{Gpaper1} consider the case $\beta = 1$ and $\beta = 3$ and find that for the two corresponding profiles $f_{\rm pin}(\rho)$ the maximum is shifted at different densities, even if the parameter $\beta$ is, of course, only a scaling factor on the same pairing gap profile. The precise height of the maximum thus depends on the vortex tension used in the model (although the order of magnitude remains $10^{15}\,\mbox{dyn}\,\mbox{cm}^{-1}$) and does not affect the location of the maximum (once $\beta$ is fixed). In this work we therefore constrain the exact value of the maximum amplitude of the pinning force by fitting the average waiting time between giant glitches in the Vela pulsar, as will be explained in the next sections. In figure \ref{fig:pinning} we show the two pinning profiles $f_{\rm pin}(\rho)$ used in this work for $\beta = 1$ and $\beta = 3$ (plotted here with the choice of $\fpm = 10^{15}\,\mbox{dyn}\,\mbox{cm}^{-1}$). 
The case $\beta_1$ has a maximum at $\rho \approx 0.325\rho_0$, while in the $\beta_3$ case the maximum is at $\rho \approx 0.14\rho_0$. In both configurations we take the pinning force to vanish at $\rho_{\rm core}$ and $\rho_d$, due to the fact that the lattice exists only in the crust and that in the outer crust there are no free neutrons to produce vortices. 

A single vortex line will be described parallel to the rotational axis and distant from this axis by a distance $x$, that represent the cylindrical radius. We consider also the vortex line to be continuous throughout the core: there is, in fact, no theoretical evidence for the existence of an interface of normal matter between the core and the inner crust, that can justify the hypothesis of a core with vorticity separated from the crust \citep{Zhou}.
Naturally the vortices may not be straight and parallel to the rotational axis, as turbulence may develop in the stellar interior, especially in the presence of strong pinning \citep{Link11,Link11b}. We will not consider this possibility here, but will discuss some of its likely consequences in the following.

\begin{figure}
	\centering
	\includegraphics[width=7cm]{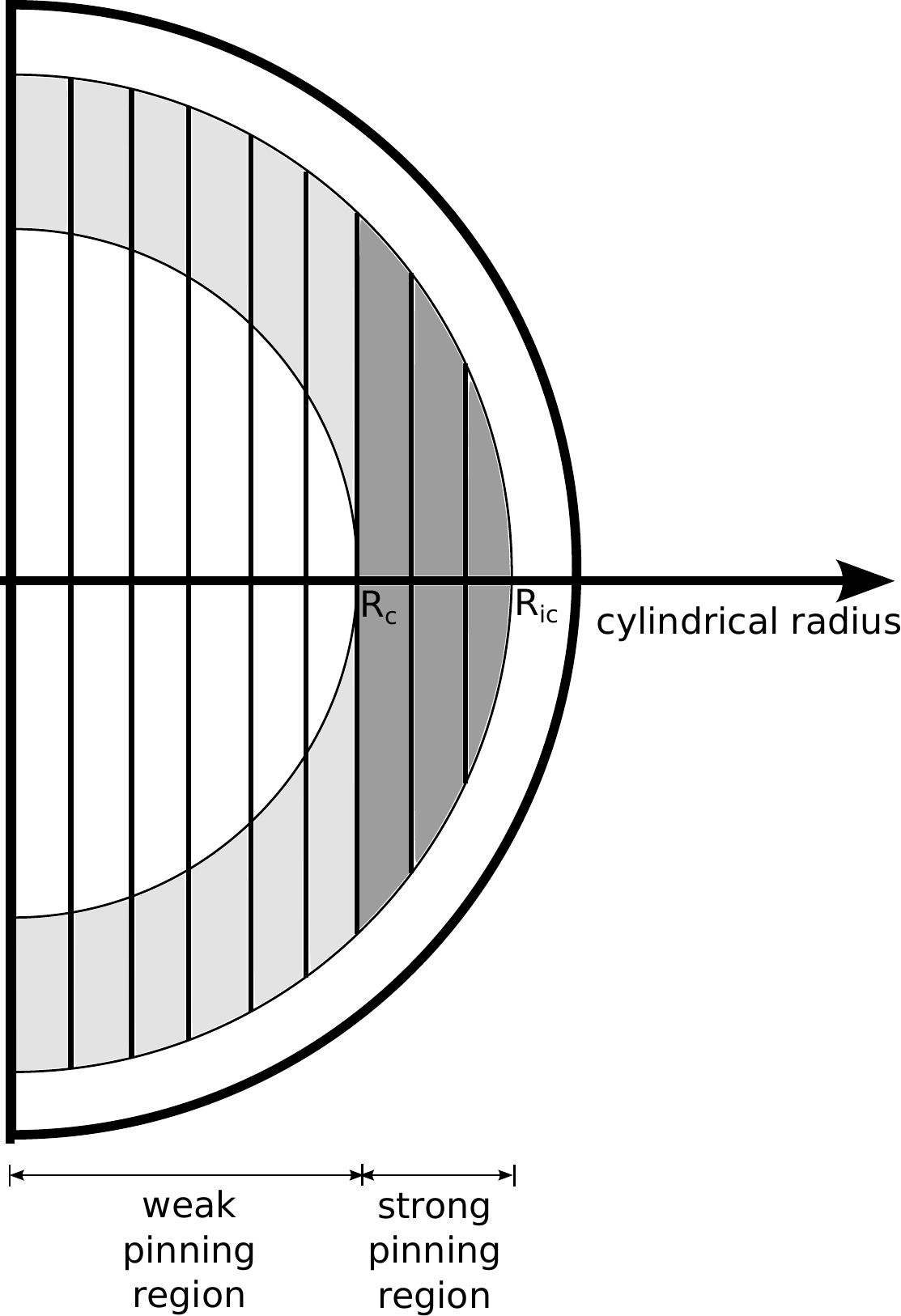}
	\caption{A schematic representation of the geometry of our problem (out of scale). The whole shaded area represents the inner crust of the NS, where vortices are pinned to the lattice. The darker part indicates the strong pinning region, where the vortices are subjected to pinning for their whole length. The star is threaded by straight continuous vortices.}
	\label{fig:schema}
\end{figure}

With the above hypothesis, we can identify two (cylindrical) pinning regions  based on the strength of the pinning interaction. The {\em strong} pinning region is  defined by $x > R_c$ and corresponds to the part of the star in which the vortices lie entirely in the inner crust region, and are therefore subject to pinning for their whole length. On the other hand, in the {\em weak} pinning region ($x < R_c$), a vortex line  is pinned only at its extremities that are immersed in the crust, while there is no pinning interaction in the core (see figure \ref{fig:schema}).

\section{The model}
Thanks to the axial symmetry of the problem, we can describe the macroscopic quantities of the rotating superfluid in terms of the variable $n(x)$ that represents the number of vortices per unit area, at a distance $x$ from the rotational axis of the star. The angular velocity $\Os(x)$ of the superfluid component of the star is in fact proportional to the number $N(x)$ of vortices enclosed in a cylindrical region of cylindrical radius $x$, and can be expressed as:
\begin{equation}
\Os(x) = \frac{\kappa}{2\pi}\frac{N(x)}{x^2} = \frac{\kappa}{2\pi x^2}\int_x n(x')\,da'
\label{eq:os}
\end{equation}
where the integration is performed on the area enclosed by the radius $x$. This result follows from the quantization of the circulation per vortex line that is encoded in the constant $\kappa = \pi \hbar / m_N$.

\begin{figure}
	\centering
	\includegraphics{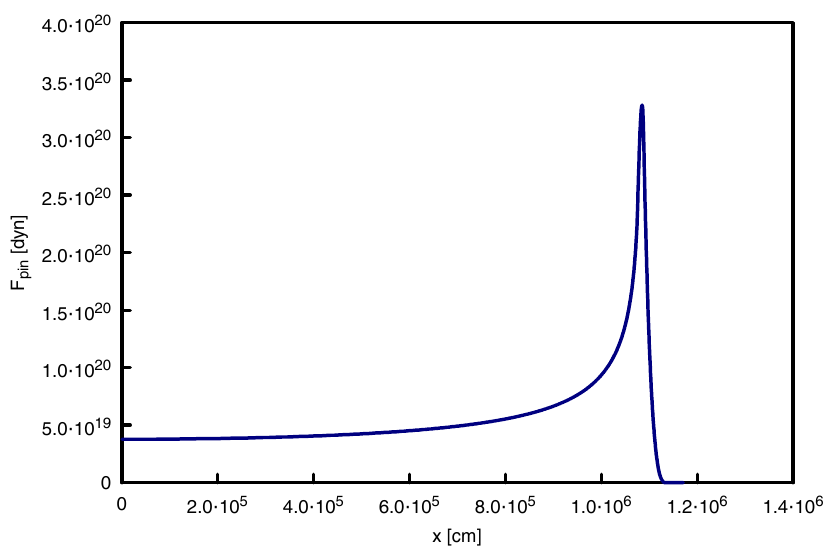}
	\caption{The total pinning force $F_{\rm pin}(x)$ integrated on the whole length of a vortex line distant $x$ from the rotational axis of the star. This plot is obtained taking a star of $1.4M_\odot$ and SLy EoS. The pinning profile used is the $\beta = 1$ case plotted in figure \ref{fig:pinning}.}
	\label{fig:totalpinning}
\end{figure}

Once a star has been fixed by the choice of an EoS and the integration of the TOV equations, the model requires, as a first step, the evaluation of the pinning force for the whole length of a generic vortex line. This can be obtained starting from the function $f_{\rm pin}(\rho)$ discussed previously. Let us imagine a vortex line parallel to the rotational axis of the star and distant $x$: the total pinning acting on it is given by the integration of $f_{\rm pin}(\rho)$ over its length:
\begin{equation}
F_{\rm pin} (x) = 2\int_0^{\ell(x)/2} f_{\rm pin}\left[ \rho\left(\sqrt{x^2 + z^2} \right)\right] \, dz
\end{equation}
where $\ell(x) = 2 \sqrt{R_{\rm ic}^2 - x^2}$ is the length of the vortex line, obtained considering that the vortex line ends at the inner--outer crust surface. To understand better the role of the pinning force, we choose a neutron star of $1.4M_\odot$ with SLy equation of state and we plot the function $F_{\rm pin}(x)$ for $x$ from $0$ to $R_{\rm ic}$ (fig. \ref{fig:totalpinning}, corresponding to $\beta = 1$ and $\fpm = 10^{15}\,\mbox{dyn}\,\mbox{cm}^{-1}$).

The pinning interaction is not the only force that acts on a vortex line. As shown in \citet{RudSut}, pinning prevents the vortex line from moving with the local superfluid velocity because the vortex line is compelled to have the velocity of the normal matter component (the normal component rotates as a rigid body with angular velocity $\Oc$). This fact give rise to a Magnus force:
\begin{equation}
\mathbf{f}_{m} = \kappa \rho_s \, \mathbf{e}_z \times (\mathbf{v}_v - \mathbf{v}_s)
\label{eq:magnus}
\end{equation}
where $\mathbf{v}_v$ is the velocity of the vortex line and $\mathbf{v}_s$ is the superfluid velocity; here $\mathbf{f}_{m}$ must be intended as force per unit length. 

In this expression $\rho_s$ is the density of the superfluid fraction of the star. In fact the whole star can be divided in two components: the normal one (which includes also the protons in the core as they are coupled with the crust by the magnetic field) and the superfluid one, on which the Magnus force will act. It thus follows that $\rho_s = (1 - x_p) \rho$ where $x_p$ is the proton fraction at a given density. Of course this quantity is a microphysical property of matter and for this reason is strictly dependent on the EoS used. As this information is not provided with the EoSs used, we use the results of \cite{Zuo} who give the proton fraction $x_p(\rho)$ as a function of the total density in the case of two--body interactions and also in the case of three-body forces. We use both the $x_p(\rho)$ relations of \cite{Zuo} but we consider also a third case where the proton fraction is a constant that does not depend on the total density. We also introduce the parameter $Q$ that represent the superfluid fraction of the star. It is defined for the general case as
\begin{equation}
Q = I_s / I_{\rm tot} = \frac{\int_0^R r^4 (1-x_p(\rho))\rho(r)\,dr}{\int_0^R r^4 \rho(r) \,dr}
\label{eq:Q}
\end{equation}
where we have used eq. \ref{eq:inertia}; $I_{\rm tot}$ is the total moment of inertia  and $I_s$ is the moment of inertia of the superfluid component. In the case of a constant proton fraction it then follows that $Q = 1 - x_p$. The average value is $Q \approx 0.95$ and therefore we shall test our model also with this prescription.

The Magnus force in \ref{eq:magnus} has only one component in the radial direction ($\mathbf{v}_v$ and $\mathbf{v}_s$ are, in fact, directed along $\mathbf{e}_\theta$, so the cross product is directed along $\mathbf{e}_x$) and therefore can be rewritten as:
\begin{equation}
\mathbf{f}_{m}(x, z) = f_m(x, z)\,\mathbf{e}_x = - \kappa \rho_s(x,z) x \DO(x) \, \mathbf{e}_x
\end{equation}
where the difference of the two velocities is written as $\Delta v(x)=x \DO(x) = x \left[ \Oc - \Os(x)\right]$ and depends only on the coordinate $x$, as described by equation \ref{eq:os}. This quantity is negative between two glitches because the normal component spins slower that the superfluid one; indeed the Magnus force is a hydrodynamical lift that pushes the vortex outward from the rotational axis. The key point here is the fact that the normal component spins down as a consequence of the loss of energy  by electromagnetic radiation of the star: the result is an increase of $f_m(x,z)$ in the time interval between glitches.

The same integration performed with $f_{\rm pin}$ over the length of the vortex can be done with the Magnus force. We can consider the total Magnus force acting on a vortex line distant $x$ from the rotational axis:
\begin{eqnarray}
F_m(x)  & = &  2 \int_0^{\ell(x)/2} f_m(x,z)\,dz \nonumber \\
& = &  2\kappa x \DO(x) \int_0^{\ell(x)/2} \rho_s \left(\sqrt{x^2 + z^2} \right) \, dz.
\end{eqnarray}

\begin{figure}
	\centering
	\includegraphics{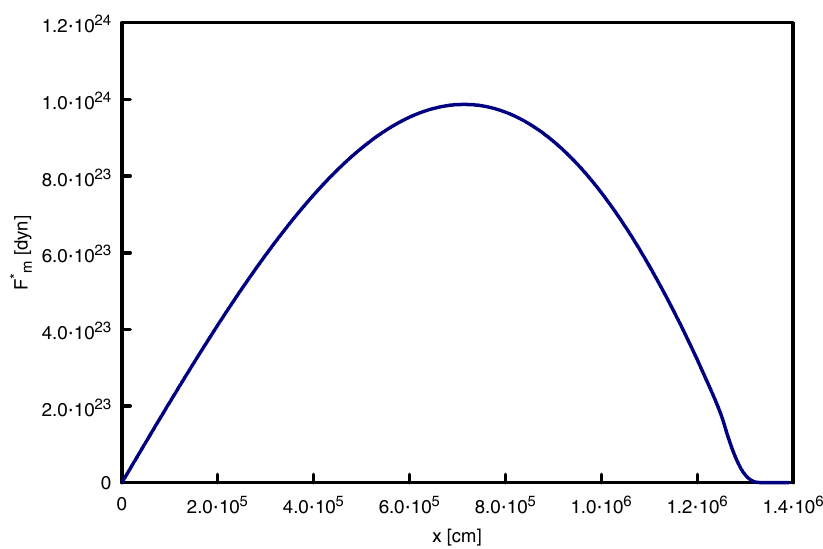}
	\caption{Plot of the expression $F^*_m(x) = F_m(x) / \DO(x)$ as a function of the cylindrical radius $x$. This plot is obtained taking a star of $1.4M_\odot$ and SLy EoS.}
	\label{fig:magnus}
\end{figure}

The basic idea here is to compare the pinning force and the Magnus force to find the critical lag $\DOC(x)$ that represents the depinning condition: when the actual lag between the two components of the stars reaches the value $\DOC$ at some point with cylindrical radius $x$, the vortices here are unbound from the lattice, as the Magnus force now exceeds the pinning interaction that held the vortices in place:
\begin{equation}
F_{\rm pin}(x) = F_m(x) = \DOC(x) F^*_m(x) 
\end{equation}
Here $F^*_m(x) = F_m(x) / \DOC(x)$ and it is plotted in fig. \ref{fig:magnus} using the same reference star as in fig. \ref{fig:totalpinning}.
The important quantity is therefore the critical lag that can be easily evaluated as:
\begin{equation}
\DOC(x) = \frac{\int_0^{\ell(x)/2} f_{\rm pin}\left[ \rho\left(\sqrt{x^2 + z^2} \right)\right] \, dz}{\kappa x  \int_0^{\ell(x)/2} \rho_s \left(\sqrt{x^2 + z^2} \right) \, dz}.
\end{equation}

\begin{figure*}
	\centering
	\includegraphics{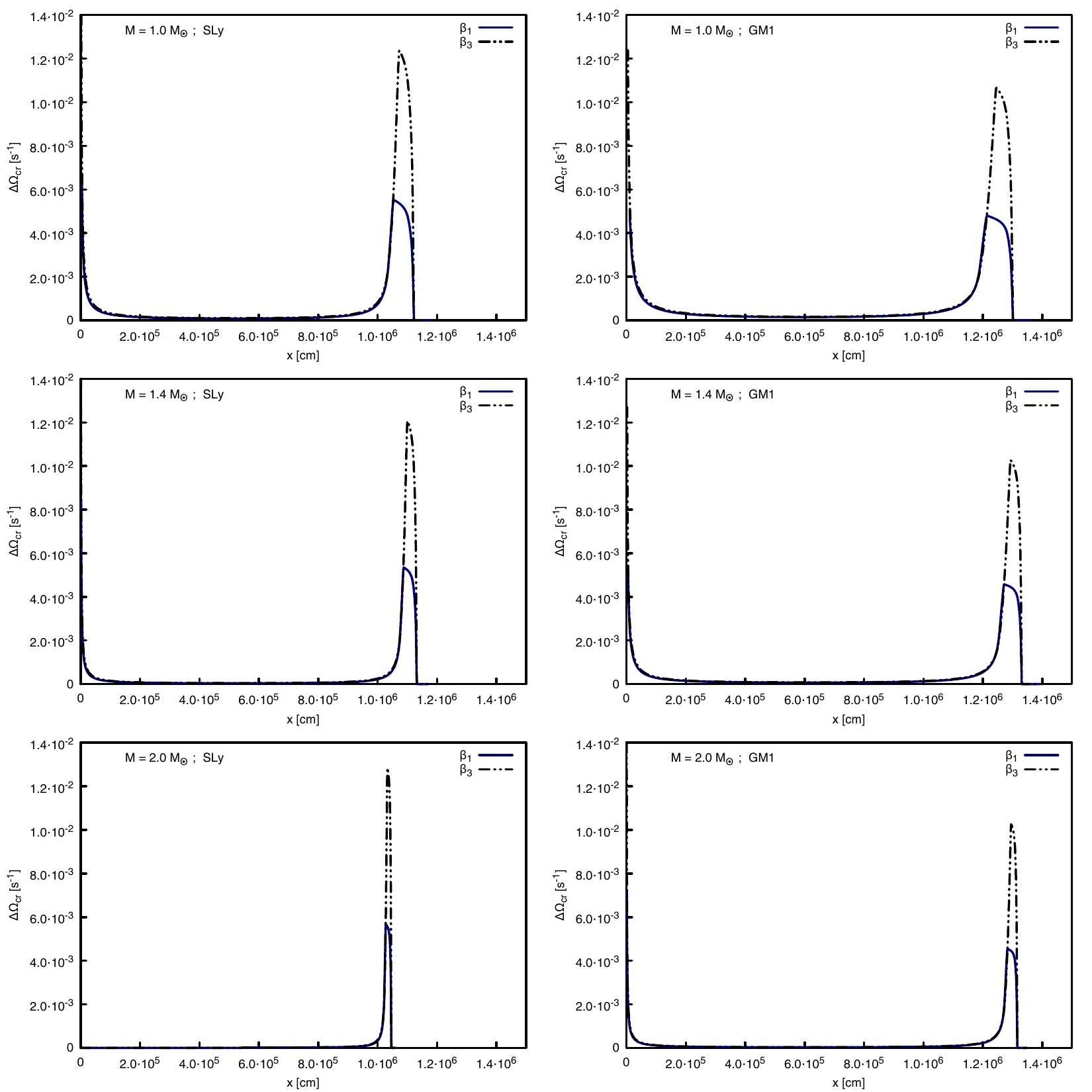}
	\caption{Plot of the critical lag $\DOC(x)$ for different stellar models, with varying mass and equations of state. The pinning profiles used are plotted in figure \ref{fig:pinning} and we consider both the case $\beta=1$ and the case $\beta=3$. Note that in both cases ($\beta=1$ and $\beta=3$) the maximum amplitude of the pinning force is the same, so the difference in the maximum lag for the two cases is now entirely due to the different position in density of the maximum in the pinning force profile.}
	\label{fig:criticallag}
\end{figure*}

In figure \ref{fig:criticallag} we plot the critical lag for sample stars from table~\ref{tab:stars}.

It is important to point out that the critical lag shows a peak $\DOCM = \DOC(\xm)$ in a region that corresponds to the inner crust, that is the region where the pinning is stronger. In this region our estimate of $\DOC(x)$ is reasonable since pinning is continuous along the whole single vortex. This is not the case for the critical lag in the core, because here pinning acts on vortices only at their extremities: as explained by \citet{Pizzochero} this fact is responsible of the {\em weak} pinning in this region, even though we can assume that the system maintains axial symmetry due to the collective rigidity of vortex bundles \citep{RudSut}. 

As the star slows down, the depinning condition $\Delta\Omega(x) \ge \DOC(x)$  is first reached in the {\em core}: as shown by \citet{Linkn}, in this region repinning is dynamically possible if the lag falls below a critical value (smaller than the one for depinning). This suggest the following interpretation: in the {\em core}, as the star slows down, the vortices are continuously depinned and repinned, establishing a dynamical creep that removes the excess vorticity on short timescales. Furthermore the Magnus force in the interior is likely to overcome the tension of vortices and depin them long before the unpinning condition in the crust is met \citep{Glaberson,Hask12}. The conclusion is thus that vortices in the core can essentially be considered free. In this region the scattering of electrons off magnetised vortex cores is mainly responsible for the drag forces and for the short relaxation timescale $\tau_c \sim 1-10 \mbox{s}$ \citep{ALS,AndSid}: this means that we can consider the normal and the superfluid components in the core as coupled with a lag of order $|\Od|\tau_c$.

In the time between glitches, the depinning region becomes larger, involving also the crust: in the inner crust the excess vorticity is repinned and creates a thin vortex sheet that moves toward the peak: this sheet is pushed outward by the increasing Magnus force and it stores angular momentum. When the peak is reached, there is no more pinning interaction that can block the excess vorticity: this vorticity is suddenly released and reaches the outer crust. At this moment the angular momentum stored by vortices is transferred to the normal component of the star, and this causes the glitch.

It is straightforward now to evaluate the time interval between two glitches: this is given by the time needed to create a lag $\DOCM$
\begin{equation}
 \DTG = \frac{\DOCM}{|\Od|}
 \label{eq:tgl}
\end{equation}
where $\Od$ is the deceleration of the normal component referred to the pre--glitch steady--state condition.

The above arguments indicates that, immediately  before a glitch, a lag of $\DOCM$ will have been created for $x = \xm$. This means that we can use equation 
\ref{eq:os} to express the number of vortices stored at the peak in the sheet just before a glitch:
\begin{equation}
\Nv = \frac{2\pi}{\kappa}  \xm^2 \DOCM.
\label{eq:nv}
\end{equation}
Due to the particular shape of the critical lag in figure \ref{fig:criticallag}, we can assume that in this moment the excess vorticity in the region $x > \xm$ has been entirely removed by the Magnus force, and therefore the $\Nv$ vortices are the only ones responsible for the transfer of angular momentum to the normal component of the star. To evaluate the angular momentum transfer we start from the definition $dL = \Os(x)\,dI_s$. As we are interested in the angular momentum stored by $\Nv$ vortices at the peak of the pinning potential, we use the relation in \ref{eq:nv} and perform the integration on the cylindrical region $\xm < x < R_c$ to obtain the requested quantity (the integration on the coordinate $x$ stops at $R_c$ due to the fact that in the outer crust there is no superfluid component):
\begin{equation}
\DLg = 2\kappa \Nv \int_{\xm}^{R_c} x \, dx \int_0^{\ell(x)} \rho_s(\sqrt{x^2 + z^2}) \, dz.
\label{eq:dlg}
\end{equation} 

Following the arguments above, at the moment of a glitch only a fraction of the core superfluidity is coupled to the normal component of the star: in fact, the rise time of a glitch ($\tau_{\rm gl}$)  is very short  and only the instantaneously depinned fraction of vorticity in the core can respond to the variation of the angular velocity of the crust. We introduce the parameter $\Ygl$ to encode this fractional quantity.
In fact the best observational upper limit is of $\tau_{\rm gl} < 40\mbox{ s}$ for the Vela 2000 glitch \citep{Dod}, while an interesting lower limit of $\tau_{\rm gl}> 10^{-4}$ ms can be set by the non--detection of gravitational waves from the Vela 2006 glitch \citep{Andrewnew}. Theoretical estimates suggest that $\tau_{\rm gl}\approx 1-10$ s \citep{Hask12}.

In the pre--glitch steady--state condition, due to the long timescales involved, we can assume $Y_\infty = 1$; but during a glitch this quantity cannot be calculated with the snowplow model as it depends on the detailed short--time dynamics of the vortices, and must thus be determined with hydrodynamical simulations such as those in \citet{Hask12}. As this is beyond the scope of the current paper, the quantity $\Ygl$ is taken as a parameter of this model, and must be inferred from the observational data as shown in the next section.

The value $\Ygl$ is needed for the evaluation of $\DOgl$, the jump in angular velocity of the normal component of the star due to a glitch. This corresponds to the ratio between the angular momentum transfer $\DLg$ and the effective moment of inertia $I_{\rm eff}$ of the coupled fraction of matter during the glitch. One thus has that $I_{\rm eff} = (1 - Q) I_{\rm tot} + Q \Ygl I_{\rm tot}$ and the requested quantity is therefore:
\begin{equation}
\DOgl = \frac{\DLg}{I_{\rm tot} \left[ 1 - Q \left( 1 - \Ygl \right) \right]}.
\label{eq:DOgl}
\end{equation}
A further parameter of the glitch that can be calculated is the relative acceleration of the crust. As illustrated in \citet{Pizzochero} the desired relation follows from variation at the glitch of the Euler equation for the normal component and angular momentum conservation:
\begin{equation}
\ACC = \frac{Q(1 - \Ygl)}{1 - Q(1 - \Ygl)}.
\end{equation}

\begin{table*}
 \centering
 \begin{minipage}{170mm}
 \caption{This table gives the fitting parameters $\fpm$ (maximum of the pinning force per unit length) and $\Ygl$ (fraction of coupled vorticity at the glitch) for all the considered configurations. $\xm$ is the position (cylindrical radius) of the maximum critical lag. These values refer to a constant proton fraction $x_p(\rho) = 0.05$ (the corresponding fraction of moment of inertia due to the superfluid component of the star is $Q = 0.95$). Unphysical (negative) values for $\Ygl$ are not reported (see text for details). Since the angular momentum transferred to the crust during a glitch is strongly dependant on the ratio $\xm/R_{\rm ic}$ (see section \ref{sec:results}), this quantity is also reported in table.}
 \label{tab:A095} 
 \setlength{\tabcolsep}{9.8pt}
\begin{tabular}{@{}lrrrrrcrrrr@{}}
\hline
& & & \multicolumn{3}{c}{$\beta = 1$} & & \multicolumn{3}{c}{$\beta = 3$} \\
\cline{3-6} \cline{8-11} \\[-2ex]
{\bf EoS} & $M$ & $\xm$ & $\xRat$ & $\fpm/10^{15}$ & $\Ygl$ & & $\xm$ & $\xRat$ & $\fpm/10^{15}$ & $\Ygl$ \\
& ($M_\odot$) & (km) & & (dyn$\,$cm$^{-1}$) & & & (km) & & (dyn$\,$cm$^{-1}$) & \\
\hline
{\bf SLy} &    1.0 &   10.530 &    0.938 &    1.562 &    0.203 & &   10.724 &    0.955 &    0.697 &    0.027\\
{\bf    } &    1.1 &   10.654 &    0.945 &    1.581 &    0.148 & &   10.829 &    0.960 &    0.704 &    0.009\\
{\bf    } &    1.2 &   10.753 &    0.951 &    1.595 &    0.106 & &   10.910 &    0.965 &    0.709 &       --\\
{\bf    } &    1.3 &   10.827 &    0.956 &    1.606 &    0.074 & &   10.968 &    0.968 &    0.713 &       --\\
{\bf    } &    1.4 &   10.875 &    0.961 &    1.613 &    0.048 & &   11.001 &    0.972 &    0.715 &       --\\
{\bf    } &    1.5 &   10.897 &    0.965 &    1.616 &    0.027 & &   11.010 &    0.975 &    0.716 &       --\\
{\bf    } &    1.6 &   10.889 &    0.969 &    1.615 &    0.011 & &   10.990 &    0.977 &    0.715 &       --\\
{\bf    } &    1.7 &   10.847 &    0.972 &    1.609 &       -- & &   10.937 &    0.980 &    0.711 &       --\\
{\bf    } &    1.8 &   10.759 &    0.975 &    1.596 &       -- & &   10.838 &    0.982 &    0.705 &       --\\
{\bf    } &    1.9 &   10.600 &    0.979 &    1.572 &       -- & &   10.667 &    0.985 &    0.694 &       --\\
{\bf    } &    2.0 &   10.279 &    0.982 &    1.525 &       -- & &   10.332 &    0.987 &    0.672 &       --\\
\hline
{\bf GM1} &    1.0 &   12.129 &    0.932 &    1.798 &    0.493 & &   12.447 &    0.956 &    0.809 &    0.078\\
{\bf    } &    1.1 &   12.315 &    0.939 &    1.825 &    0.389 & &   12.604 &    0.961 &    0.819 &    0.051\\
{\bf    } &    1.2 &   12.473 &    0.945 &    1.849 &    0.307 & &   12.737 &    0.965 &    0.827 &    0.031\\
{\bf    } &    1.3 &   12.604 &    0.950 &    1.868 &    0.242 & &   12.844 &    0.968 &    0.834 &    0.015\\
{\bf    } &    1.4 &   12.710 &    0.955 &    1.884 &    0.190 & &   12.929 &    0.971 &    0.840 &    0.003\\
{\bf    } &    1.5 &   12.792 &    0.959 &    1.896 &    0.148 & &   12.992 &    0.974 &    0.844 &       --\\
{\bf    } &    1.6 &   12.852 &    0.962 &    1.905 &    0.113 & &   13.034 &    0.976 &    0.847 &       --\\
{\bf    } &    1.7 &   12.890 &    0.966 &    1.910 &    0.083 & &   13.055 &    0.978 &    0.848 &       --\\
{\bf    } &    1.8 &   12.901 &    0.969 &    1.912 &    0.060 & &   13.052 &    0.980 &    0.848 &       --\\
{\bf    } &    1.9 &   12.885 &    0.972 &    1.910 &    0.040 & &   13.022 &    0.982 &    0.846 &       --\\
{\bf    } &    2.0 &   12.836 &    0.974 &    1.902 &    0.023 & &   12.960 &    0.984 &    0.842 &       --\\
{\bf    } &    2.1 &   12.744 &    0.977 &    1.889 &    0.007 & &   12.854 &    0.985 &    0.835 &       --\\
{\bf    } &    2.2 &   12.586 &    0.980 &    1.865 &       -- & &   12.683 &    0.987 &    0.824 &       --\\
{\bf    } &    2.3 &   12.287 &    0.982 &    1.821 &       -- & &   12.367 &    0.989 &    0.803 &       --\\
\hline
\end{tabular}
 \end{minipage}
\end{table*}

\begin{table*}
 \centering
 \begin{minipage}{170mm}
 \caption{The fitting parameters given here (defined in table \ref{tab:A095}) refer to the proton fraction $x_p(\rho)$ proposed by \citet{Zuo}, obtained with two--body forces.}
 \label{tab:AZU1} 
 \setlength{\tabcolsep}{9.8pt}
\begin{tabular}{@{}lrrrrrcrrrr@{}}
\hline
& & & \multicolumn{3}{c}{$\beta = 1$} & & \multicolumn{3}{c}{$\beta = 3$} \\
\cline{3-6} \cline{8-11} \\[-2ex]
{\bf EoS} & $M$ & $\xm$ & $\xRat$ & $\fpm/10^{15}$ & $\Ygl$ & & $\xm$ & $\xRat$ & $\fpm/10^{15}$ & $\Ygl$ \\
& ($M_\odot$) & (km) & & (dyn$\,$cm$^{-1}$) & & & (km) & & (dyn$\,$cm$^{-1}$) & \\
\hline
{\bf SLy} &    1.0 &   10.530 &    0.938 &    1.618 &    0.212 & &   10.724 &    0.955 &    0.727 &    0.029\\
{\bf    } &    1.1 &   10.654 &    0.945 &    1.637 &    0.152 & &   10.829 &    0.960 &    0.735 &    0.007\\
{\bf    } &    1.2 &   10.753 &    0.951 &    1.652 &    0.105 & &   10.910 &    0.965 &    0.740 &       --\\
{\bf    } &    1.3 &   10.827 &    0.956 &    1.663 &    0.067 & &   10.968 &    0.968 &    0.744 &       --\\
{\bf    } &    1.4 &   10.875 &    0.961 &    1.671 &    0.036 & &   11.001 &    0.972 &    0.746 &       --\\
{\bf    } &    1.5 &   10.897 &    0.965 &    1.674 &    0.010 & &   11.010 &    0.975 &    0.747 &       --\\
{\bf    } &    1.6 &   10.889 &    0.969 &    1.673 &       -- & &   10.990 &    0.977 &    0.745 &       --\\
{\bf    } &    1.7 &   10.847 &    0.972 &    1.666 &       -- & &   10.937 &    0.980 &    0.742 &       --\\
{\bf    } &    1.8 &   10.759 &    0.975 &    1.653 &       -- & &   10.837 &    0.982 &    0.735 &       --\\
{\bf    } &    1.9 &   10.600 &    0.979 &    1.628 &       -- & &   10.666 &    0.985 &    0.723 &       --\\
{\bf    } &    2.0 &   10.279 &    0.982 &    1.579 &       -- & &   10.332 &    0.987 &    0.701 &       --\\
\hline
{\bf GM1} &    1.0 &   12.129 &    0.932 &    1.861 &    0.522 & &   12.446 &    0.956 &    0.843 &    0.099\\
{\bf    } &    1.1 &   12.315 &    0.939 &    1.890 &    0.415 & &   12.605 &    0.961 &    0.854 &    0.070\\
{\bf    } &    1.2 &   12.473 &    0.945 &    1.914 &    0.330 & &   12.736 &    0.965 &    0.863 &    0.048\\
{\bf    } &    1.3 &   12.604 &    0.950 &    1.934 &    0.262 & &   12.844 &    0.968 &    0.870 &    0.029\\
{\bf    } &    1.4 &   12.710 &    0.955 &    1.951 &    0.207 & &   12.929 &    0.971 &    0.876 &    0.014\\
{\bf    } &    1.5 &   12.792 &    0.959 &    1.963 &    0.162 & &   12.992 &    0.974 &    0.880 &    0.002\\
{\bf    } &    1.6 &   12.852 &    0.962 &    1.972 &    0.124 & &   13.034 &    0.976 &    0.883 &       --\\
{\bf    } &    1.7 &   12.888 &    0.966 &    1.978 &    0.092 & &   13.055 &    0.978 &    0.884 &       --\\
{\bf    } &    1.8 &   12.900 &    0.969 &    1.980 &    0.064 & &   13.052 &    0.980 &    0.884 &       --\\
{\bf    } &    1.9 &   12.885 &    0.972 &    1.977 &    0.039 & &   13.022 &    0.982 &    0.882 &       --\\
{\bf    } &    2.0 &   12.836 &    0.974 &    1.970 &    0.018 & &   12.960 &    0.984 &    0.878 &       --\\
{\bf    } &    2.1 &   12.744 &    0.977 &    1.956 &       -- & &   12.854 &    0.985 &    0.871 &       --\\
{\bf    } &    2.2 &   12.586 &    0.980 &    1.931 &       -- & &   12.683 &    0.987 &    0.859 &       --\\
{\bf    } &    2.3 &   12.286 &    0.982 &    1.885 &       -- & &   12.367 &    0.989 &    0.838 &       --\\
\hline
\end{tabular}
 \end{minipage}
\end{table*}

\begin{table*}
 \centering
 \begin{minipage}{170mm}
 \caption{This table is analogous to tables \ref{tab:A095} and \ref{tab:AZU1}: here the proton fraction used is that calculated by \citet{Zuo} with three--body forces.}
 \label{tab:AZU2} 
 \setlength{\tabcolsep}{9.8pt}
\begin{tabular}{@{}lrrrrrcrrrr@{}}
\hline
& & & \multicolumn{3}{c}{$\beta = 1$} & & \multicolumn{3}{c}{$\beta = 3$} \\
\cline{3-6} \cline{8-11} \\[-2ex]
{\bf EoS} & $M$ & $\xm$ & $\xRat$ & $\fpm/10^{15}$ & $\Ygl$ & & $\xm$ & $\xRat$ & $\fpm/10^{15}$ & $\Ygl$ \\
& ($M_\odot$) & (km) & & (dyn$\,$cm$^{-1}$) & & & (km) & & (dyn$\,$cm$^{-1}$) & \\
\hline
{\bf SLy} &    1.0 &   10.530 &    0.938 &    1.619 &    0.197 & &   10.724 &    0.955 &    0.728 &    0.011\\
{\bf    } &    1.1 &   10.654 &    0.945 &    1.638 &    0.133 & &   10.829 &    0.960 &    0.735 &       --\\
{\bf    } &    1.2 &   10.753 &    0.951 &    1.653 &    0.080 & &   10.910 &    0.965 &    0.740 &       --\\
{\bf    } &    1.3 &   10.827 &    0.956 &    1.664 &    0.036 & &   10.968 &    0.968 &    0.744 &       --\\
{\bf    } &    1.4 &   10.875 &    0.961 &    1.672 &       -- & &   11.001 &    0.972 &    0.746 &       --\\
{\bf    } &    1.5 &   10.897 &    0.965 &    1.675 &       -- & &   11.010 &    0.975 &    0.747 &       --\\
{\bf    } &    1.6 &   10.889 &    0.969 &    1.674 &       -- & &   10.990 &    0.977 &    0.746 &       --\\
{\bf    } &    1.7 &   10.847 &    0.972 &    1.667 &       -- & &   10.937 &    0.980 &    0.742 &       --\\
{\bf    } &    1.8 &   10.759 &    0.975 &    1.654 &       -- & &   10.837 &    0.982 &    0.735 &       --\\
{\bf    } &    1.9 &   10.600 &    0.979 &    1.629 &       -- & &   10.666 &    0.985 &    0.724 &       --\\
{\bf    } &    2.0 &   10.279 &    0.982 &    1.580 &       -- & &   10.332 &    0.987 &    0.701 &       --\\
\hline
{\bf GM1} &    1.0 &   12.129 &    0.932 &    1.862 &    0.521 & &   12.446 &    0.956 &    0.843 &    0.095\\
{\bf    } &    1.1 &   12.315 &    0.939 &    1.891 &    0.413 & &   12.605 &    0.961 &    0.854 &    0.065\\
{\bf    } &    1.2 &   12.473 &    0.945 &    1.915 &    0.326 & &   12.736 &    0.965 &    0.863 &    0.042\\
{\bf    } &    1.3 &   12.604 &    0.950 &    1.935 &    0.257 & &   12.844 &    0.968 &    0.870 &    0.022\\
{\bf    } &    1.4 &   12.710 &    0.955 &    1.952 &    0.200 & &   12.929 &    0.971 &    0.876 &    0.006\\
{\bf    } &    1.5 &   12.792 &    0.959 &    1.964 &    0.153 & &   12.992 &    0.974 &    0.880 &       --\\
{\bf    } &    1.6 &   12.852 &    0.962 &    1.973 &    0.112 & &   13.034 &    0.976 &    0.883 &       --\\
{\bf    } &    1.7 &   12.888 &    0.966 &    1.979 &    0.077 & &   13.055 &    0.978 &    0.885 &       --\\
{\bf    } &    1.8 &   12.900 &    0.969 &    1.981 &    0.046 & &   13.052 &    0.980 &    0.884 &       --\\
{\bf    } &    1.9 &   12.885 &    0.972 &    1.978 &    0.017 & &   13.022 &    0.982 &    0.882 &       --\\
{\bf    } &    2.0 &   12.836 &    0.974 &    1.971 &       -- & &   12.960 &    0.984 &    0.878 &       --\\
{\bf    } &    2.1 &   12.744 &    0.977 &    1.957 &       -- & &   12.854 &    0.985 &    0.871 &       --\\
{\bf    } &    2.2 &   12.586 &    0.980 &    1.932 &       -- & &   12.683 &    0.987 &    0.859 &       --\\
{\bf    } &    2.3 &   12.286 &    0.982 &    1.886 &       -- & &   12.367 &    0.989 &    0.838 &       --\\
\hline
\end{tabular}
 \end{minipage}
\end{table*}

\begin{table*}
 \centering
 \begin{minipage}{170mm}
 \caption{This table shows, for the considered configurations, the physical quantities that the ``snowplow'' model permits to evaluate: the number $\Nv$ of vortices stored at the peak in critical lag just before the glitch, the angular momentum transferred to the crust $\DLg$, and the step in frequency derivative on short timescales $\Acc$. Like table \ref{tab:A095} (that gives the fitting parameters used), this one refers to a constant proton fraction that gives $Q = 0.95$.}
 \label{tab:B095} 
 \setlength{\tabcolsep}{11.6pt}
\begin{tabular}{@{}lrrrrrcrrr@{}}
\hline
& & & \multicolumn{3}{c}{$\beta = 1$} & & \multicolumn{3}{c}{$\beta = 3$} \\
\cline{4-6} \cline{8-10} \\[-2ex]
{\bf EoS} & $M$ & $R$ & $\Nv$ & $\DLg$ & $\Acc$ & & $\Nv$ & $\DLg$ & $\Acc$ \\
& ($M_\odot$) & (km) & ($10^{13}$) & ($10^{40}$ erg$\,$s) & & & ($10^{13}$) & ($10^{40}$ erg$\,$s) & \\
\hline
{\bf SLy} &    1.0 &   11.855 &    3.041 &    3.889 &    3.124 & &    3.154 &    1.216 &   12.184\\
{\bf    } &    1.1 &   11.830 &    3.113 &    3.425 &    4.242 & &    3.216 &    1.059 &   15.950\\
{\bf    } &    1.2 &   11.797 &    3.171 &    2.996 &    5.622 & &    3.265 &    0.919 &       --\\
{\bf    } &    1.3 &   11.758 &    3.215 &    2.601 &    7.331 & &    3.299 &    0.792 &       --\\
{\bf    } &    1.4 &   11.705 &    3.244 &    2.235 &    9.476 & &    3.319 &    0.677 &       --\\
{\bf    } &    1.5 &   11.635 &    3.257 &    1.903 &   12.166 & &    3.325 &    0.570 &       --\\
{\bf    } &    1.6 &   11.545 &    3.252 &    1.595 &   15.629 & &    3.313 &    0.477 &       --\\
{\bf    } &    1.7 &   11.422 &    3.227 &    1.304 &       -- & &    3.280 &    0.387 &       --\\
{\bf    } &    1.8 &   11.260 &    3.175 &    1.033 &       -- & &    3.221 &    0.304 &       --\\
{\bf    } &    1.9 &   11.025 &    3.082 &    0.771 &       -- & &    3.121 &    0.226 &       --\\
{\bf    } &    2.0 &   10.620 &    2.898 &    0.498 &       -- & &    2.928 &    0.147 &       --\\
\hline
{\bf GM1} &    1.0 &   13.940 &    4.034 &   11.480 &    0.931 & &    4.249 &    2.741 &    7.086\\
{\bf    } &    1.1 &   13.943 &    4.159 &   10.433 &    1.384 & &    4.357 &    2.456 &    9.128\\
{\bf    } &    1.2 &   13.940 &    4.267 &    9.429 &    1.927 & &    4.449 &    2.180 &   11.657\\
{\bf    } &    1.3 &   13.930 &    4.357 &    8.481 &    2.570 & &    4.524 &    1.958 &   14.462\\
{\bf    } &    1.4 &   13.913 &    4.430 &    7.599 &    3.330 & &    4.584 &    1.743 &   17.882\\
{\bf    } &    1.5 &   13.885 &    4.488 &    6.764 &    4.242 & &    4.629 &    1.542 &       --\\
{\bf    } &    1.6 &   13.845 &    4.530 &    5.977 &    5.344 & &    4.660 &    1.355 &       --\\
{\bf    } &    1.7 &   13.788 &    4.557 &    5.198 &    6.744 & &    4.674 &    1.184 &       --\\
{\bf    } &    1.8 &   13.715 &    4.565 &    4.543 &    8.336 & &    4.672 &    1.020 &       --\\
{\bf    } &    1.9 &   13.620 &    4.553 &    3.895 &   10.385 & &    4.651 &    0.869 &       --\\
{\bf    } &    2.0 &   13.495 &    4.519 &    3.284 &   12.999 & &    4.606 &    0.727 &       --\\
{\bf    } &    2.1 &   13.330 &    4.455 &    2.688 &   16.541 & &    4.532 &    0.593 &       --\\
{\bf    } &    2.2 &   13.095 &    4.345 &    2.104 &       -- & &    4.411 &    0.460 &       --\\
{\bf    } &    2.3 &   12.713 &    4.140 &    1.481 &       -- & &    4.195 &    0.327 &       --\\
\hline
\end{tabular}
 \end{minipage}
\end{table*}

\begin{table*}
 \centering
 \begin{minipage}{170mm}
 \caption{The quantities $\DLg$ and $\Acc$ here reported follows from calculation base on the proton fraction proposed by \citet{Zuo} with two--body interactions. The corresponding fitting parameters are shown in table \ref{tab:AZU1}. $Q = I_s / I_{\rm tot}$ is the global superfluid fraction of moment of inertia.}
 \label{tab:BZU1} 
 \setlength{\tabcolsep}{9.5pt}
\begin{tabular}{@{}lrrrrrrcrrr@{}}
\hline
& & & & \multicolumn{3}{c}{$\beta = 1$} & & \multicolumn{3}{c}{$\beta = 3$} \\
\cline{5-7} \cline{9-11} \\[-2ex]
{\bf EoS} & $M$ & $R$ & $Q$ & $\Nv$ & $\DLg$ & $\Acc$ & & $\Nv$ & $\DLg$ & $\Acc$ \\
& ($M_\odot$) & (km) & & ($10^{13}$) & ($10^{40}$ erg$\,$s) & & & ($10^{13}$) & ($10^{40}$ erg$\,$s) & \\
\hline
{\bf SLy} &    1.0 &   11.855 &    0.948 &    3.041 &    4.048 &    2.962 & &    3.154 &    1.272 &   11.609\\
{\bf    } &    1.1 &   11.830 &    0.945 &    3.113 &    3.565 &    4.036 & &    3.216 &    1.108 &   15.210\\
{\bf    } &    1.2 &   11.797 &    0.942 &    3.171 &    3.119 &    5.362 & &    3.265 &    0.961 &       --\\
{\bf    } &    1.3 &   11.758 &    0.938 &    3.215 &    2.708 &    7.005 & &    3.299 &    0.828 &       --\\
{\bf    } &    1.4 &   11.705 &    0.934 &    3.244 &    2.327 &    9.065 & &    3.319 &    0.708 &       --\\
{\bf    } &    1.5 &   11.635 &    0.930 &    3.257 &    1.981 &   11.650 & &    3.325 &    0.596 &       --\\
{\bf    } &    1.6 &   11.545 &    0.925 &    3.252 &    1.660 &       -- & &    3.313 &    0.499 &       --\\
{\bf    } &    1.7 &   11.422 &    0.920 &    3.227 &    1.357 &       -- & &    3.280 &    0.405 &       --\\
{\bf    } &    1.8 &   11.260 &    0.913 &    3.175 &    1.076 &       -- & &    3.221 &    0.321 &       --\\
{\bf    } &    1.9 &   11.025 &    0.905 &    3.082 &    0.803 &       -- & &    3.120 &    0.239 &       --\\
{\bf    } &    2.0 &   10.620 &    0.890 &    2.898 &    0.518 &       -- & &    2.928 &    0.153 &       --\\
\hline
{\bf GM1} &    1.0 &   13.940 &    0.966 &    4.034 &   11.939 &    0.856 & &    4.248 &    2.874 &    6.713\\
{\bf    } &    1.1 &   13.943 &    0.964 &    4.159 &   10.850 &    1.292 & &    4.357 &    2.559 &    8.719\\
{\bf    } &    1.2 &   13.940 &    0.962 &    4.267 &    9.805 &    1.814 & &    4.448 &    2.303 &   10.982\\
{\bf    } &    1.3 &   13.930 &    0.961 &    4.357 &    8.820 &    2.433 & &    4.524 &    2.047 &   13.792\\
{\bf    } &    1.4 &   13.913 &    0.959 &    4.430 &    7.903 &    3.164 & &    4.584 &    1.822 &   17.063\\
{\bf    } &    1.5 &   13.885 &    0.956 &    4.488 &    7.034 &    4.041 & &    4.629 &    1.612 &   20.994\\
{\bf    } &    1.6 &   13.845 &    0.954 &    4.530 &    6.216 &    5.101 & &    4.660 &    1.416 &       --\\
{\bf    } &    1.7 &   13.788 &    0.952 &    4.556 &    5.463 &    6.367 & &    4.674 &    1.237 &       --\\
{\bf    } &    1.8 &   13.715 &    0.949 &    4.564 &    4.743 &    7.943 & &    4.672 &    1.066 &       --\\
{\bf    } &    1.9 &   13.620 &    0.946 &    4.553 &    4.051 &    9.948 & &    4.651 &    0.909 &       --\\
{\bf    } &    2.0 &   13.495 &    0.942 &    4.519 &    3.415 &   12.463 & &    4.606 &    0.760 &       --\\
{\bf    } &    2.1 &   13.330 &    0.938 &    4.455 &    2.796 &       -- & &    4.532 &    0.620 &       --\\
{\bf    } &    2.2 &   13.095 &    0.933 &    4.345 &    2.188 &       -- & &    4.411 &    0.481 &       --\\
{\bf    } &    2.3 &   12.713 &    0.925 &    4.140 &    1.552 &       -- & &    4.195 &    0.342 &       --\\
\hline
\end{tabular}
 \end{minipage}
\end{table*}

\begin{table*}
 \centering
 \begin{minipage}{170mm}
 \caption{The proton fraction used for this table is the  three--body forces model of \citet{Zuo}. The corresponding fitting parameters are shown in table \ref{tab:AZU2}.}
 \label{tab:BZU2} 
 \setlength{\tabcolsep}{9.5pt}
\begin{tabular}{@{}lrrrrrrcrrr@{}}
\hline
& & & & \multicolumn{3}{c}{$\beta = 1$} & & \multicolumn{3}{c}{$\beta = 3$} \\
\cline{5-7} \cline{9-11} \\[-2ex]
{\bf EoS} & $M$ & $R$ & $Q$ & $\Nv$ & $\DLg$ & $\Acc$ & & $\Nv$ & $\DLg$ & $\Acc$ \\
& ($M_\odot$) & (km) & & ($10^{13}$) & ($10^{40}$ erg$\,$s) & & & ($10^{13}$) & ($10^{40}$ erg$\,$s) & \\
\hline
{\bf SLy} &    1.0 &   11.855 &    0.931 &    3.041 &    4.049 &    2.960 & &    3.154 &    1.272 &   11.607\\
{\bf    } &    1.1 &   11.830 &    0.924 &    3.113 &    3.567 &    4.034 & &    3.216 &    1.108 &       --\\
{\bf    } &    1.2 &   11.797 &    0.916 &    3.171 &    3.120 &    5.360 & &    3.265 &    0.961 &       --\\
{\bf    } &    1.3 &   11.758 &    0.908 &    3.215 &    2.708 &    7.002 & &    3.299 &    0.828 &       --\\
{\bf    } &    1.4 &   11.705 &    0.899 &    3.244 &    2.328 &       -- & &    3.319 &    0.708 &       --\\
{\bf    } &    1.5 &   11.635 &    0.888 &    3.257 &    1.981 &       -- & &    3.325 &    0.597 &       --\\
{\bf    } &    1.6 &   11.545 &    0.876 &    3.252 &    1.661 &       -- & &    3.313 &    0.499 &       --\\
{\bf    } &    1.7 &   11.422 &    0.862 &    3.227 &    1.358 &       -- & &    3.280 &    0.405 &       --\\
{\bf    } &    1.8 &   11.260 &    0.844 &    3.175 &    1.076 &       -- & &    3.221 &    0.321 &       --\\
{\bf    } &    1.9 &   11.025 &    0.820 &    3.082 &    0.803 &       -- & &    3.120 &    0.239 &       --\\
{\bf    } &    2.0 &   10.620 &    0.779 &    2.898 &    0.518 &       -- & &    2.928 &    0.153 &       --\\
\hline
{\bf GM1} &    1.0 &   13.940 &    0.962 &    4.034 &   11.944 &    0.856 & &    4.248 &    2.874 &    6.712\\
{\bf    } &    1.1 &   13.943 &    0.959 &    4.159 &   10.854 &    1.292 & &    4.357 &    2.559 &    8.718\\
{\bf    } &    1.2 &   13.940 &    0.957 &    4.267 &    9.809 &    1.813 & &    4.448 &    2.303 &   10.981\\
{\bf    } &    1.3 &   13.930 &    0.953 &    4.357 &    8.824 &    2.432 & &    4.524 &    2.047 &   13.791\\
{\bf    } &    1.4 &   13.913 &    0.950 &    4.430 &    7.906 &    3.163 & &    4.584 &    1.822 &   17.061\\
{\bf    } &    1.5 &   13.885 &    0.946 &    4.488 &    7.037 &    4.039 & &    4.629 &    1.612 &       --\\
{\bf    } &    1.6 &   13.845 &    0.942 &    4.530 &    6.218 &    5.099 & &    4.660 &    1.416 &       --\\
{\bf    } &    1.7 &   13.788 &    0.937 &    4.556 &    5.465 &    6.365 & &    4.674 &    1.237 &       --\\
{\bf    } &    1.8 &   13.715 &    0.931 &    4.564 &    4.744 &    7.939 & &    4.672 &    1.066 &       --\\
{\bf    } &    1.9 &   13.620 &    0.925 &    4.553 &    4.052 &    9.944 & &    4.651 &    0.909 &       --\\
{\bf    } &    2.0 &   13.495 &    0.917 &    4.519 &    3.416 &       -- & &    4.606 &    0.760 &       --\\
{\bf    } &    2.1 &   13.330 &    0.907 &    4.455 &    2.797 &       -- & &    4.532 &    0.620 &       --\\
{\bf    } &    2.2 &   13.095 &    0.894 &    4.345 &    2.188 &       -- & &    4.411 &    0.481 &       --\\
{\bf    } &    2.3 &   12.713 &    0.873 &    4.140 &    1.553 &       -- & &    4.195 &    0.342 &       --\\
\hline
\end{tabular}
 \end{minipage}
\end{table*}

\section{Results and observations}
\label{sec:results}
In this section we test the model proposed here against observations. As the model has been developed for giant glitches we shall compare our results to observations of giant glitches in the Vela pulsar. The Vela (PSR B0833-45 or PSR J0835-4510) has a spin frequency $\nu\approx 11.19$ Hz and spin-down rate $\dot{\nu}\approx -1.55 \times 10^{-11}$ Hz s$^{-1}$; from relation \ref{eq:tgl} this value correspond to a maximum critical lag of $\DOCM = 8.6\times 10^{-3} \mbox{ rad}\,\mbox{s}^{-1}$, where we have considered that the average time between glitches for this pulsar is $2.8$ years. The glitch is usually described in terms of permanent steps in the frequency and frequency derivative and a series of transient terms that decay exponentially. It is well known that at least three transient terms are required, with decay timescales that range from months to hours \citep{Flanagan}. Recent observations of the 2000 and 2004 glitch have shown that an additional term is required on short timescales, with a decay time of approximately a minute.
Given that the detection in 2004 was only barely above the noise we shall refer to the January 2000 glitch. In this case the jump in angular velocity was of $\DOgl = 2.2 \times 10^{-4} \mbox{ rad}\,\mbox{s}^{-1}$ \citep{Dod,dodson2}. This is a fairly typical value for giant glitches in the Vela, and we take it as our reference value. The relative step in frequency derivative corresponding to the transient term with the shortest decay timescale (1 minute) for this glitch is $\Acc \approx 18 \pm 6$ (1$\sigma$ error), and we assume that this is a good approximation to the {\em instantaneous}  step in frequency derivative at the time of the glitch.

As explained in the previous section, the model has two free parameters that are the maximum value of the pinning force value $\fpm$ and $\Ygl$: this means that, once a star has been fixed (by choosing the EoS, the mass $M$, and the superfluid fraction relation) we can use two observational quantities  to constrain the parameters of the model  and compare further observables to the quantities predicted by calculations. In particular, for each fixed star, we rescale the maximum of the pinning force in order to produce the maximum critical lag $\DOCM$ required to reproduce the average waiting time between glitches in the Vela. This allows  us to calculate directly and univocally the angular momentum $\DLg$ from equation \ref{eq:dlg}. As we want to reproduce a glitch of amplitude $\DOgl = 2.2 \times 10^{-4} \mbox{ rad}\,\mbox{s}^{-1}$, relation \ref{eq:DOgl} can be rewritten in the following form 
\begin{equation}
\Ygl = \frac{1}{Q}\left[\frac{\DLg}{I_{\rm tot} \DOgl} + Q -1  \right],
\end{equation}
and therefore can be used to fix the coupled fraction of matter during the glitch. Tables \ref{tab:A095}, \ref{tab:AZU1} and \ref{tab:AZU2} give the fitting parameters for all  the configurations tested.
We can see that the value of the maximum pinning force $\fpm$ does not change significantly with the total mass of the star.
In these tables negative values of $\Ygl$ are not given as they would not be physically acceptable: a negative value would mean that there is not enough angular momentum to produce the required jump in angular velocity, even if we consider the core vorticity completely decoupled from the normal component of the star at the time of the glitch.

The remaining tables, numbered \ref{tab:B095}, \ref{tab:BZU1} and \ref{tab:BZU2}, show the physical quantities that the ``snowplow'' model permits to evaluate. These are of course the angular momentum $\DLg$ transferred to the crust during the glitch and the relative step in frequency derivative. We can see that the order of magnitude for $\DLg$ is $10^{40}\,\mbox{erg}\,\mbox{s}$, that is compatible with the upper limits on the glitch energy obtained from observations of the power wind nebula surrounding Vela \citep{Helfand} and with the results found in \citet{Pizzochero}, where the same model is applied analytically with a polytropic EoS in Newtonian gravity. From these tables one can see that, for a particular choice of EoS and proton fraction, the angular momentum $\DLg$ stored by vortices decreases with the total mass of the star.
This behaviour can be easily explained, as shown by \citet{Pizzochero}, in terms of the quantity $\xm/R_{\rm ic}$ shown in the tables: $\DLg$ is obviously related to the number $\Nv$ of vortices stored at the peak (in tables \ref{tab:B095}, \ref{tab:BZU1} and \ref{tab:BZU2}; see also eq. \ref{eq:dlg}) -- that however doesn't change significantly with the mass -- but it depends strongly on the ratio $\xm/R_{\rm ic}$ which increases at higher masses. In \citet{Pizzochero} (fig. 4) it is clearly shown that the angular momentum stored by the vortices at the peak decreases rapidly moving the position of the peak towards the outer crust. The quantity $\DLg$ also depends on the equation of state used (a stiffer EoS produces higher values of $\DLg$) and on the pinning profile: the $\beta= 3$ condition, when other variables are fixed, gives lower values for the angular momentum, accordingly to the fact that the relative position of the peak with respect to the inner crust radius is higher.

The ``snowplow'' model permits to calculate also the step in spin--down rate immediately after a glitch, and this quantity is given in our tables as $\Acc$. It has been calculated only for acceptable values of $\Ygl$, and must be compared with the reference value of $\Acc = 18 \pm 6$, taken from the Vela 2000 glitch \citep{Dod}. These values suggest that the $\beta = 3$ configurations are preferred and this can be considered in reasonable agreement with the microscopic results found by \citet{Gandolfi}: they find that a realistic suppression factor for the pairing gap $\Delta(\rho)$ is $\beta \approx 1.5$ but, crucially, also that the maximum for $\Delta(\rho)$ is shifted at lower densities.  This leads to a profile close to what we obtain for $\beta=3$ in our model.

Finally let us remark that the results in tables \ref{tab:B095}, \ref{tab:BZU1} and \ref{tab:BZU2}, for the (microscopically favoured) case $\beta=3$, seem to indicate that a stiffer equation of state (GM1) is preferred as is a lower mass (possibly in the region of $1.4 M_\odot$) for the Vela pulsar. Naturally such a quantitative conclusion is difficult to make on the basis of one observation and it would be highly desirable to have information on the short-timescale post-glitch behaviour not only of other Vela giant glitches, but also of other glitching pulsars.
Note that short term components of the relaxation have not been measured for other giant glitchers, however the ``snowplow'' model can be used to predict waiting times, obtaining results which are consistent with observations \citep{Hask12}.

\section{Conclusions}
In this paper we have extended the ``snowplow'' model for giant pulsar glitches of \citet{Pizzochero} to incorporate relativistic background stellar models and realistic equations of state. In particular we test the model for the SLy and GM1 equations of state. Unfortunately these equations of state do not include information on beta equilibrium, so we use the proton fractions calculated by \citet{Zuo}. It would of course be highly desirable to use proton fractions that are consistent with the individual equations of state in future work, in order to set stringent constraints. Furthermore we use, for the first time, the realistic profiles for the pinning forces per unit length calculated by \citet{Gpaper1} (see also \citet{Gtesi,Gpaper2}), in order to evaluate the amount of angular momentum that can be transferred to the crust during a glitch. 

The model contains three free parameters, the mass of the star $M$, the fraction of superfluid that is coupled to the crust during a glitch $Y_{gl}$, (which can only be estimated with dynamical simulations such as those of \citet{Hask12}), and the maximum amplitude of the pinning force, $\fpm$. Note in fact that while the location of the maximum is precisely determined by the microphysical calculations of \citet{Gpaper1} (see also \citet{Gtesi,Gpaper2}), the actual value of the maximum can vary by factors of order unity or more as it depends on the poorly constrained value of the vortex tension. We thus treat it as a normalization and determine its value by requiring that the waiting time between glitches is of 2.8 years, as is approximately the case for Vela glitches.  We then fit the size of the glitch to an average Vela glitch  to obtain the value of $Y_{gl}$. In particular we take the value of the Vela 2000 glitch, $\Delta\Omega=2.2\times10^{-4}\mbox{ rad}\,\mbox{s}^{-1}$ \citep{Dod}.

Having determined the free parameters in our model, except for the mass of the NS which is free, we compare our results to the post glitch step in frequency derivative. Unfortunately the changes in $\dot{\nu}$ on short time scales after a glitch are observationally challenging to detect and it has been possible to fit for transient steps in frequency and frequency derivative on timescales of minutes after a glitch only for the Vela 2000 and 2004 glitch \citep{Dod,dodson2}. Given that the detection is only barely above the noise for the 2004 glitch \citep{dodson2} we fit to the values obtained for the 2000 glitch, which we assume to be a good approximation of the instantaneous post glitch behaviour. This justifies our choice of also fitting to the value of the jump in frequency of the Vela 2000 glitch.

The comparison of the model to the observational constraints first of all highlights that the general results of the analytic model of \citet{Pizzochero} remain valid even in our more physically realistic approach and the results are in general consistent for both equations of state for a reasonable range of neutron star masses. The glitch model presented here thus appears robust and compatible with the observations of giant glitches in the Vela and is, as shown in \citet{Hask12}, compatible with the average waiting time between giant glitches in other pulsars.  This further reinforces the hypothesis that giant glitches are approximately periodic phenomena that occur close to the maximum lag that the pinning force can support in the crust, while smaller glitches may be triggered by random events such as crust quakes \citep{Rud76,Rud} or vortex avalanches \citep{Melatos1,Melatos2,Melatos3,Andrewnew}.
Furthermore our results favour lower masses for the Vela pulsar (smaller than $1.5M_{\odot}$) and stiffer equations of state. Note however that such a quantitative conclusion is difficult to draw as not only are we comparing to a single observation but dynamical simulations have also shown that superfluid mutual friction will contribute significantly to the short term post-glitch spindown \citep{Hask12} as may friction at the crust/core interface \citep{vE1}. In will thus be necessary to further develop hydrodynamical glitch simulations to truly constrain the equation of state and the stellar mass.

Finally let us remark that in this paper we have assumed straight vortices that cross cross the core of the neutron star and are only weakly pinned at their extremities. Although the assumption of vortices that pass through the star appears to be justified by microphysical estimates, that do not predict an interface of normal matter between the crust and core superfluid \citep{Zhou}, it may be the case that if the protons in the core are in a type II superconducting state this could lead to strong pinning also in the stellar interior \citep{Rud,Link03}.
In this case not only would vortex motion be impeded, but it is also likely that turbulence will develop \citep{Link11,Link11b}.  Note, however, that a large portion of the star may be in a type I superconducting state \citep{Jones06} in which the magnetic field is not organised in flux tubes, but rather in macroscopic regions of normal matter, and the interactions may be much weaker \citep{Sedrakian05} (although see \citet{Jones06} for a discussion of pinning in type I superconductors). Furthermore \citet{Babaev} has recently shown that in the presence of strong entrainment or superfluid $\Sigma^{-}$ hyperons, the interaction between vortices and flux tubes may be much weaker than generally assumed in the presence of type II superconductivity. In this paper we thus take the view that pinning in the core will be weak, although strong pinning of vortices to flux tubes is an intriguing possibility and will be the focus of a future publication (Haskell, Pizzochero \& Seveso, in preparation).

Turbulence, on the other hand, is well known from laboratory superfluids and may play an important role in pulsar glitches \citep{Peralta06,Peralta07,Peralta09} and could couple the superfluid and the normal component on inter-glitch timescales in the presence of core pinning \citep{Trev07}. The inclusion of turbulence in a hydrodynamical glitch simulations is, however, a complex matter as not only is the nature of the turbulence not known (see e.g. \citet{Trev07,Link11,Link11b}) but also the definition of pinning force per unit length must be revisited in the presence of a turbulent tangle. Such a fundamental issue should clearly be the focus of future work.

\section*{Acknowledgments}

This work was supported by CompStar, a Research Networking Programme of the European Science Foundation.
BH acknowledges support from the European Union via a Marie-Curie IEF fellowship
and from the European Science Foundation (ESF) for the activity entitled ``The New Physics of Compact Stars'' (COMPSTAR) under exchange grant 2449.

\nocite{*}
\bibliographystyle{mn2e}

\begin{thebibliography}{99}
\bibitem[\protect\citeauthoryear{Adams Cieplak \& Glaberson}{Adams, Cieplak \& Glaberson}{1984}]{Glaberson} Adams P.W., Cieplak M., Glaberson W.I., 1984., Phys Rev B, 32, 171
\bibitem[\protect\citeauthoryear{Alpar}{1977}]{Alpar77} Alpar M.A., 1977, Ap.J. 213, 527
\bibitem[\protect\citeauthoryear{Alpar et al.}{1984a}]{Alpar84} Alpar M.A., Anderson P.W., Pines D., Shaham J., 1984, Ap.J. 278, 791
\bibitem[\protect\citeauthoryear{Alpar et al.}{1984b}]{ALS} Alpar M.A., Langer S.A., Sauls J.A., 1984, Ap.J. 282, 533
\bibitem[\protect\citeauthoryear{Anderson \& Itoh}{1975}]{AndItoh75} Anderson P.W., Itoh N., 1975, Nature 256, 25
\bibitem[\protect\citeauthoryear{Anderson et al.}{1982}]{Anderson82} Anderson P.W., Alpar M.A., Pines D., Shaham J., 1982, Philos.Mag. A 45, 227
\bibitem[\protect\citeauthoryear{Andersson, Sidery \& Comer}{2006}]{AndSid} Andersson N., Sidery T., Comer G.L., 2006, MNRAS 368, 162
\bibitem[\protect\citeauthoryear{Andersson, Sidery \& Comer}{2007}]{Trev07} Andersson N., Sidery T., Comer G.L., 2007, MNRAS 381, 747
\bibitem[\protect\citeauthoryear{Babaev}{2009}]{Babaev} Babaev E., 2009, Phys.Rev.Lett 103, 231101
\bibitem[\protect\citeauthoryear{Baym, Pathick \& Pines}{1969}]{Baym} Baym G., Pethick C., Pines D., 1969, Nature 224, 872
\bibitem[\protect\citeauthoryear{Crawford \& Demianski}{2003}]{Crow} Crawford F., Demianski M., 2003, ApJ 595, 1052
\bibitem[\protect\citeauthoryear{Demorest et al.}{2010}]{Dem} Demorest P.B., Pennucci T., Ransom S.M., Roberts M.S.E., Hessels J.W.T., 2010, Nature 467, 1081
\bibitem[\protect\citeauthoryear{Dodson, McCulloch \& Lewis}{2002}]{Dod} Dodson R.G., McCulloch P.M., Lewis D.R., 2002, ApJL 564, L85
\bibitem[\protect\citeauthoryear{Dodson, Lewis \& McCulloch}{Dodson et al.}{2007}]{dodson2} Dodson R.G., Lewis D.R.,  McCulloch P.M., 2007, Ap\&SS, 308, 585
\bibitem[\protect\citeauthoryear{Donati \& Pizzochero}{Donati \& Pizzochero}{2003}]{DP03} Donati, P., Pizzochero P.M., 2003, Phys.Rev.Lett. 90, 21
\bibitem[\protect\citeauthoryear{Donati \& Pizzochero}{Donati \& Pizzochero}{2004}]{DP04} Donati, P., Pizzochero P.M., 2004, Nu.Phys.A, 742, 363
\bibitem[\protect\citeauthoryear{Donati \& Pizzochero}{Donati \& Pizzochero}{2006}]{DP06} Donati, P., Pizzochero P.M., 2006, Phys.Lett.B, 640

\bibitem[\protect\citeauthoryear{Douchin \& Haensel}{2001}]{DH01} Douchin F., Haensel P., 2001, A\&A 380, 151
\bibitem[\protect\citeauthoryear{Epstein \& Baym}{Epstein \& Baym}{1988}]{EB} Epstein R.I., Baym G., 1988, Ap.J. 328, 680
\bibitem[\protect\citeauthoryear{van Eysden \& Melatos}{van Eysden \& Melatos}{2010}]{vE1} van Eysden C.A., Melatos A., 2010, MNRAS 409, 1253
\bibitem[\protect\citeauthoryear{Espinoza, Lyne, Stappers \& Kramer}{Espinoza et al.}{2011}]{Espinoza} Espinoza C.M., Lyne A.G., Stappers B.W, Kramer M.,  2011, MNRAS 414, 1679
\bibitem[\protect\citeauthoryear{Flanagan}{Flanagan}{1996}]{Flanagan} Flanagan C.S., 1996, ``Pulsars: Problems \& Progress'', ASP Conference Series, Vol. 105, Eds. S.Johnston, M.A. Walker \& M.Bailes

\bibitem[\protect\citeauthoryear{Gandolfi, Illarionov, Fantoni, Pederiva, \& Schmidt}{Gandolfi et al.}{2008}]{Gandolfi} Gandolfi S., Illarionov A.Yu, Fantoni S., Pederiva F., Schmidt K.E., 2008., Phys Rev Lett., 101, 132501

\bibitem[\protect\citeauthoryear{Glampedakis \& Andersson}{Glampedakis \& Andersson}{2009}]{kglitch} Glampedakis K., Andersson N., 2009., Phys Rev Lett., 102, 141101
\bibitem[\protect\citeauthoryear{Glampedakis \& Andersson}{2011a}]{Glamp11} Glampedakis K., Andersson N., 2011a, ApJ 740, L35

\bibitem[\protect\citeauthoryear{Glendenning \& Moszokowski}{1991}]{GM1} Glendenning, N.K., Moszkowski, S.A. 1991, Phys. Rev. Lett. 67, 2414
\bibitem[\protect\citeauthoryear{Grill}{2011}]{Gtesi} Grill F., 2011, PhD Thesis, University of Milan.
\bibitem[\protect\citeauthoryear{Grill \& Pizzochero}{2012a}]{Gpaper1} Grill F., Pizzochero P.M., 2012a, Journal of Physics: Conference Series 342, 012004
\bibitem[\protect\citeauthoryear{Grill \& Pizzochero}{2012b}]{Gpaper2} Grill F., Pizzochero P.M., 2012b, in preparation

\bibitem[\protect\citeauthoryear{Haskell, Pizzochero \& Sidery}{2012}]{Hask12} Haskell B., Pizzochero P.M., Sidery T., 2012, MNRAS 420, 658
\bibitem[\protect\citeauthoryear{Helfand et al.}{2001}]{Helfand} Helfand D. J., Gotthelf E. V., \& Halper, J. P. 2001, ApJ, 556, 380

\bibitem[\protect\citeauthoryear{Jones}{2006}]{Jones06} Jones P.B., 2006, MNRAS 371, 1327
\bibitem[\protect\citeauthoryear{Larson \& Link}{Larson \& Link}{2002}]{LL} Larson M.B., Link B., 2002, MNRAS, 333, 613
\bibitem[\protect\citeauthoryear{Link}{2003}]{Link03} Link B., 2003, Phys.Rev.Lett. 91 ,10110
\bibitem[\protect\citeauthoryear{Link}{ Link}{2009}]{Linkn} Link B., 2009, Phys.Rev.Letters 102, 131101
\bibitem[\protect\citeauthoryear{Link}{2011a}]{Link11b} Link B., 2011a, preprint: arXiv:1105.4654
\bibitem[\protect\citeauthoryear{Link}{2011b}]{Link11} Link B., 2011b, preprint: arXiv:1111.0696
\bibitem[\protect\citeauthoryear{Melatos \& Peralta}{2007}]{Peralta07} Melatos A., Peralta C., 2007, ApJ. 662, L99
\bibitem[\protect\citeauthoryear{Melatos \& Warszawski}{2009}]{Melatos2} Melatos A., Warszawski L., 2009, ApJ. 700, 1524
\bibitem[\protect\citeauthoryear{Middleditch et al.}{2006}]{Mid06} Middleditch J., Marshall F.E., Wang Q.D., Gotthelf E.V., Zhang W., 2006, ApJ, 625, 1531
\bibitem[\protect\citeauthoryear{Page et al.}{2011}]{CASA1} Page D., Prakash M., Lattimer J.M., Steiner A.W., 2011, Phys.Rev.Lett. 106, 081101
\bibitem[\protect\citeauthoryear{Peralta et al.}{2006}]{Peralta06} Peralta C., Melatos A., Giacobello M., Ooi A., 2006, ApJ. 651,
\bibitem[\protect\citeauthoryear{Peralta \& Melatos}{2009}]{Peralta09} Peralta C., Melatos A., 2009, ApJ. 701, L75
 1079
\bibitem[\protect\citeauthoryear{Pines et al.}{1980}]{Pines80} Pines D., Shaham J., Alpar M.A., Anderson P.W., 1980, Prog.Theor.Phys., Suppl. 69, 376
\bibitem[\protect\citeauthoryear{Pizzochero}{2011}]{Pizzochero} Pizzochero P.M., 2011, ApJ. 743, L20
\bibitem[\protect\citeauthoryear{Ruderman \& Sutherland}{1974}]{RudSut} Ruderman M.A., Sutherland P.G., 1974, Ap.J. 190, 137
\bibitem[\protect\citeauthoryear{Ruderman}{1976}]{Rud76} Ruderman M., 1976, Ap.J. 203, 213
\bibitem[\protect\citeauthoryear{Ruderman, Zhu \& Chen}{Ruderman, Zhu \& Chen}{1998}]{Rud} Ruderman M., Zhu T., Chen K., 1998, Ap.J. 492, 267
\bibitem[\protect\citeauthoryear{Sedrakian}{2005}]{Sedrakian05} Sedrakian A., 2005, Phys.Rev.D 71, 3003
\bibitem[\protect\citeauthoryear{Shternin et al.}{2011}]{CASA2} Shternin P.S., Yakovlev D.G., Heinke C.O., Ho W.C.G., Patnaude D.J., 2011, MNRAS 41, L108
\bibitem[\protect\citeauthoryear{Warszawski \& Melatos}{2008}]{Melatos1} Warszawski L., Melatos A., 2008, MNRAS 390, 175
\bibitem[\protect\citeauthoryear{Warszawski \& Melatos}{2011}]{Melatos3} Warszawski L., Melatos A., 2011, MNRAS 415, 1611
\bibitem[\protect\citeauthoryear{Warszawski \& Melatos}{2012}]{Andrewnew} Warszawski L., Melatos A., 2012, preprint: arXiv:1203.4466

\bibitem[\protect\citeauthoryear{Zhou et al.}{2004}]{Zhou} Zhou X.-R., Schulze H.-J., Zhao E.-G., Pan F., Draayer J. P., 2004, Phys. Rev. C, 70, 048802
\bibitem[\protect\citeauthoryear{Zuo et al.}{2004}]{Zuo} Zuo W., Li Z.H., Lu G.C., Li J.Q., Scheid W., Lombardo U., Schulze H.-J., Shen C.W., 2004, Phys.Lett.B 595, 44
\end{thebibliography}

\bsp

\label{lastpage}

\end{document}